\documentclass[]{aastex631}

\usepackage{color}

\newcommand{\btz}{\texttt{baseline\_v2.0}\ }
\newcommand{\btzns}{\texttt{baseline\_v2.0}}
\newcommand{\bto}{\texttt{baseline\_v2.1}\ }
\newcommand{\btons}{\texttt{baseline\_v2.1}}
\newcommand{\bof}{\texttt{baseline\_v1.5}\ }
\newcommand{\bofns}{\texttt{baseline\_v1.5}}

\shorttitle{LSST Survey Strategy in the Galactic Plane}
\shortauthors{Street et al.}
\graphicspath{{./}{figures/}}
\begin{document}
%\nolinenumbers

\title{LSST Survey Strategy in the Galactic Plane and Magellanic Clouds}

\author[0000-0001-6279-0552]{R.A.~Street}
\affiliation{Las Cumbres Observatory, 6740 Cortona Drive, Suite 102, Goleta, CA 93117, USA}

\author[0000-0002-0514-5650]{X.~Li}
\affiliation{Department of Physics \& Astronomy, University of Delaware, 217 Sharp Lab, Newark, DE 19716, USA}

\author[0000-0002-1910-7065]{S.~Khakpash}
\affiliation{Department of Physics \& Astronomy, University of Delaware, 217 Sharp Lab, Newark, DE 19716, USA}
\affiliation{Rutgers, The State University of New Jersey,
Department of Physics \& Astronomy, 136 Frelinghuysen Rd, Piscataway, NJ 08854, USA}

\author{E.~Bellm}
\affiliation{DiRAC Institute, Department of Astronomy, University of Washington, 3910 15th Avenue NE, Seattle, WA 98195, USA}

\author[0000-0002-6301-3269]{L.~Girardi}
\affiliation{Osservatorio Astronomico di Padova, INAF, Vicolo dell'Osservatorio 5, I-35122 Padova, Italy}

\author{L.~Jones}
\affiliation{DiRAC Institude, University of Washington, PAB Building (Physics-Astronomy Building), 3910 15th Ave NE, Seattle, WA 98195, USA}

\author[0000-0002-0287-3783]{N. S. Abrams}
\affiliation{Department of Astronomy, 501 Campbell Hall \#3411, University of California at Berkeley, Berkeley, CA 94720-3411, USA}

\author[0000-0001-8411-351X]{Y.~Tsapras}
\affiliation{Astronomisches Rechen-Institut, M{\"o}nchhofstr. 12-14, D-69120 Heidelberg,
Germany}

\author{M.P.G.~Hundertmark}
\affiliation{Astronomisches Rechen-Institut, M{\"o}nchhofstr. 12-14, D-69120 Heidelberg,
Germany}

\author[0000-0002-6578-5078]{E.~Bachelet}
\affiliation{IPAC, Mail Code 100-22, Caltech, 1200 E. California Blvd., Pasadena, CA 91125, USA}

\author{P. Gandhi}
\affiliation{University of Southampton, B46, West Highfield Campus, University Road, Southampton, SO17 1BJ, UK}

\author{P. Szkody}
\affiliation{Department of Astronomy, University of Washington, Box 351580, Seattle, WA 98195-1700, USA}

\author[0000-0002-2577-8885]{W. I. Clarkson}
\affiliation{Department of Natural Sciences, University of Michigan-Dearborn, 4901 Evergreen Road, Dearborn, MI 48128, USA}

\author[0000-0002-3258-1909]{R. Szab\'o}
\affiliation{Konkoly Observatory, CSFK, MTA Centre of Excellence, Budapest, Konkoly Thege Miklós {\'u}t 15-17. H-1121 Hungary}
\affiliation{MTA CSFK Lend\"ulet Near-Field Cosmology Research Group,  H-1121 Budapest, Konkoly Thege Mikl\'os \'ut 15-17, Hungary}
\affiliation{ELTE Eötvös Loránd University, Institute of Physics, 1117, P\'azm\'any P\'eter s\'et\'any 1/A, Budapest, Hungary}

\author{L. Prisinzano}
\affiliation{INAF - Osservatorio Astronomico di Palermo, Piazza del Parlamento, 1 90134 Palermo, Italy}

\author[0000-0001-9297-7748]{R. Bonito}
\affiliation{INAF - Osservatorio Astronomico di Palermo, Piazza del Parlamento, 1 90134 Palermo, Italy}

\author{D. A. H. Buckley}
\affiliation{South African Astronomical Observatory, PO Box 9, Observatory Rd, 7935 Observatory, Cape Town, South Africa}
\affiliation{Department of Astronomy, University of Cape Town, Private Bag X3, Rondebosch 7701, South Africa}
\affiliation{Department of Physics, University of the Free State, PO Box 339, Bloemfontein 9300, South Africa}

\author{J. P. Marais}
\affiliation{Department of Physics, University of the Free State, PO Box 339, Bloemfontein 9300, South Africa}

\author[0000-0003-0972-1376]{R. Di Stefano}
\affiliation{Harvard-Smithsonian Center for Astrophysics 60 Garden Street, Office: P-227. Cambridge, MA 02138, USA}



\begin{abstract}
Galactic science encompasses a wide range of subjects in the study of the Milky Way and Magellanic Clouds, from Young Stellar Objects to X-ray Binaries.  Mapping these populations, and exploring transient phenomena within them, are among the primary science goals of the Vera C. Rubin Observatory's Legacy Survey of Space and Time (LSST).  While early versions of the survey strategy dedicated relatively few visits to the Galactic Plane region, more recent strategies under consideration envision higher cadence within selected regions of high scientific interest. The range of galactic science presents a challenge in evaluating which strategies deliver the highest scientific returns.  Here we present metrics designed to evaluate Rubin survey strategy simulations based on the cadence of observations they deliver within regions of interest to different topics in galactic science, using variability categories defined by timescale.  We also compare the fractions of exposures obtained in each filter with those recommended for the different science goals.   We find that the {\tt baseline\_v2.x} simulations deliver observations of the high-priority regions at sufficiently high cadence to reliably detect variability on timescales $>$10\,d or more.   Follow-up observations may be necessary to properly characterize variability, especially transients, on shorter timescales.  Combining the regions of interest for all the science cases considered, we identify those areas of the Galactic Plane and Magellanic Clouds of highest priority. We recommend that these refined survey footprints be used in future simulations to explore rolling cadence scenarios, and to optimize the sequence of observations in different bandpasses.
\end{abstract}



\section{Introduction} 
\label{sec:intro}
The National Science Foundation's Vera C. Rubin Observatory Legacy Survey of Space and Time, (LSST), will cover the entire southern sky visible from the telescope's site at Cerro Pach\'{o}n, Chile, but not to uniform limiting magnitude or temporal cadence.  Figure~\ref{fig:pencilbeams_map} illustrates the number of visits realized to different regions of the sky, comparing the {\tt baseline\_v1.5} and {\tt baseline\_v2.0} iterations of the \textbf{main, Wide-Fast-Deep (WFD) survey strategy.}  Regions at lower ($|b|<15^{\circ}$) galactic latitude and extreme $\gtrsim$60$^{\circ}$ declination received significantly fewer visits than the rest of the sky over the 10-yr span of LSST in {\tt baseline\_v1.5}.  The distribution of visits was adjusted in {\tt baseline\_v2.0} to include the Magellenic Clouds and central Bulge, though much of the Galactic Plane still receives relatively few exposures.

This survey strategy places severe limitations on the science that the survey will realize, particularly for stars and planets in the Milky Way.  Eleven of the 46 White Papers on LSST survey cadence submitted in 2018\footnote{ \url{https://www.lsst.org/submitted-whitepaper-2018}} argued for increasing the number of visits to the Galactic Plane and Magellanic Clouds.  These papers identified a vast array of scientific inquiries that could be revolutionized by systematic observations from Rubin Observatory, including: 

\begin{itemize}
    \item Exploring the atmospheric properties of time-variable brown dwarfs, and extending the known population of transiting planets and white dwarfs \citep{lund2018};
    \item Comparing Pre-Main Sequence stars in Star Forming Regions (SFRs) with a range of metallicities down to the lowest mass stars; using stellar variability and photometric properties to determine the ages and distances of Open and Globular Clusters, and so informing our understanding of the evolution of the variables within those clusters \citep{prisinzano2018}; 
    \item Using the photometric variability of Young-Stellar Objects to characterize mass accretion from circumstellar disks as well as magnetic cycles and flares \citep{bonito2018}; 
    \item Developing a more complete view of the population of Cataclysmic variables and AM CVn stars, detected from their outbursts and different accretion states, to be compared with close-binary evolution models and to track the accretion behavior of white dwarfs \citep{olsen2018};
    \item Detecting and monitoring outbursts from X-ray Binaries that will reveal the population of neutron stars and black holes, determining whether a mass gap exists between them, establishing whether supernova explosions give large black hole kicks and identifying them as a potential source of gravitational wave events \citep{strader2018};
    \item Producing extensive catalogs of stellar flares that will enable them to be studied as a function of stellar properties, shedding light on the star's internal structure and processes; obtaining pulsations of white dwarfs to trace their cooling curves, and exploring pulsations as a function of metallicity; measuring the ages and metallicities of stars in the Galactic Bulge and tracing the formation history of the region \citep{gonzalez2018, bono2018}; 
    \item Using microlensing by stars and planets to reveal the distribution of these populations in different stellar environments throughout the galaxy while also discovering isolated black holes and enabling the mass function of these objects to be fully described \citep{street2018diverse}; 
    \item Using RR Lyrae stars, Cepheids, SX Phoenicis, delta Scuti stars and Long Period Variables (LPVs) within resolved stellar populations to trace different stellar generations over the large spatial extension and magnitude depth allowed by the LSST.  Measuring their distances using variable stars of different type/parent stellar population will map their 3D structures, and detect tidal streams \citep{clementini2018};
    \item Detecting recurrent novae and determining whether they are progenitors of supernovae Type Ia \citep{strader2018}; 
\end{itemize} 

Several papers highlighted LSST's ability to map the unique extra-galactic stellar populations of the Magellanic Clouds, down to faint limiting magnitudes, if they were included in the survey footprint.  LSST could therefore produce a complete catalog of variability for these populations, including detecting the microlensing signatures of planets \citep{poleski2018, olsen2018, street2018diverse}.  Using DECam imaging of crowded fields as a proxy for LSST data, \cite{suberlak2018} showed that a single, 30\,s exposure will detect, 5$\sigma$ above the background sky, all objects brighter than $r_{lim}\leq$21.5\,mag in crowded regions, and probe to even fainter magnitudes, $r_{lim}$=24.5\,mag, in uncrowded areas of the sky.  

Rubin observations of the Galactic Bulge can also enhance the scientific return of other contemporary surveys, in particular the Nancy Grace Roman Space Telescope, which will provide very high-cadence near-infrared time-series observations, with the goal of detecting planets and stars via their microlensing signatures, but only for a few periods of $\sim$72\,days each.  Rubin observations could detect planets that would otherwise be missed by Roman, and enable the mass and distances of free-floating planets to be measured \citep{street2018bulge}, if its survey of the Bulge region is coordinated with Roman, and realizes adequate cadence of imaging.  

Many of these science goals do not depend on achieving deep co-added images, in contrast to the WFD.  Rather, their observational goals require repeated visits to the Galactic Plane and Magellanic Clouds in order to achieve time-series photometric measurements.  In some cases, achieving the science goals would require a different cadence of observations in different filters, in light of the greater extinction compared with the WFD region.  

LSST must meet its original scientific requirements \citep{lsst_srd}, meaning that modifications to the survey strategy can only take up to a maximum of $\sim$10\% of the available on-sky time.  These factors combine to argue that a distinct survey strategy should be considered for the Galactic Plane and Magellanic Clouds to maximize the scientific return within the survey requirements.  The Rubin Observatory could run this survey in parallel with, or as part of, the WFD, during the same 10-yr period.  

Rubin Observatory designed the LSST Metric Analysis Framework  \citep[MAF]{jones2021} to incorporate a range of metrics, including those contributed by the community, to evaluate the impact of different survey strategies on different science cases.  Many of the papers quoted above proposed metrics specific to each science topic, such as quantifying the numbers of a particular population that LSST will discover, or determining how well a given parameter will be measured.  Detailed analysis of these science cases is presented in a number of parallel papers \citep[e.g.]{Raiteri2022}, and properly interpreting their output sometimes requires a nuanced, in-depth understanding of each science topic.  The challenge for the Rubin Observatory is to find a strategy that optimizes the survey for a wide range of science, and to this end, the Rubin Survey Cadence Optimization Committee (SCOC) requested broader Figures of Merit which could be used as a guide when comparing different strategies for quite different science goals.  Rubin Observatory have recently conducted operational simulations (or "OpSims", as described in \citealt{bianco2022}) of a wide range of alternative survey strategies, and invited community feedback on the scientific impacts of different configurations. 

The goal of this paper is to review the observational needs of a range of different galactic science topics to identify areas of overlap and tension in survey strategy, and to recommend scientifically-motivated solutions.  We present metrics designed to evaluate the essential observational parameters, so that the strengths and weaknesses of simulations of the different survey strategies may be evaluated, while representing a broad range of science goals. 

In Section~\ref{sec:survey_strategy_elements} we outline the essential components of any survey strategy, and describe our approach to capturing the requirements of galactic science for each component.  In Section~\ref{sec:eval_baseline2} we apply these metrics to compare the \btz strategy with previous iterations, and in Section~\ref{sec:eval_v2.0_opsims}, we evaluate the set of OpSim experiments conducted by Rubin in early 2022.  In Section~\ref{sec:refinedDiamond}, we combine the regions of interest for different galactic science use-cases to identify the highest priority regions to survey, and summarize our key findings in Section~\ref{sec:conclusions}.  

\section{Elements of Survey Strategy}
\label{sec:survey_strategy_elements}
In an ideal world, LSST would survey the entire visible sky with a regular cadence in all six filters.  To do so would unfortunately reduce the achievable cadence within the WFD region to the point where the LSST Science Requirements are not met.  More realistically, some trade-off is likely to be necessary between regular revisits to each telescope pointing on-sky (the cadence), the number and duration of exposures in different filters, and the on-sky area included in the survey footprint.  For this reason, it is valuable to consider the survey footprint, cadence and filter selection independently, so that metrics can evaluate the strengths of different strategies.  

Studies of the populations of the Galactic Plane and Magellanic Clouds naturally follow the non-uniform spatial distribution of those populations.  That is, galactic science needs observations in fairly well-defined regions at low ($|b| < 10^{\circ}$) galactic latitude, with additional regions of interest centered on the Magellanic Clouds, star clusters and star forming regions.  When the 2018 White Papers advocating for galactic science were compared, it was evident that many of them recommended observations of regions with a high degree of overlap.  This led us to explore whether a common footprint could be defined, encompassing the regions of highest scientific priority.  

The high extinction in much of the Galactic Plane tends to result in a preference for observations in redder passbands ($g,r,i,z$), particularly for time-series observations.  $y$ is not included in this set because declines in the telescope and instrument throughput and detector sensitivity at these wavelengths produces comparatively low signal-to-noise in this filter. Nevertheless, observations in $u$ and $y$ filters are still important for characterizing young star variability and spectral typing. The optimal total number and frequency of observations is therefore not uniform across all passbands.  

By focusing on the observational requirements, rather than science-specific parameters, we can combine the needs of a diverse range of science in a concise set of metrics, which we describe in the following sections.  For reference the metrics are summarized in Table~\ref{tab:metrics}.  All metric code has been integrated with the open-source Rubin MAF \citep{rubin_maf}.  

\begin{table}[ht]
    \centering
    \begin{tabular}{|l|l|l|l|}
    \hline
        MAF Module & MAF Metric & Metric Output & Evaluated Metrics \\
                   &            & per HEALpix & \\
    \hline
        \texttt{galacticPlaneMetric} & \texttt{GalPlaneFootprintMetric} & No. observations & \%ofPriority($\tau_{var}$,map)\\
                                   &   & nObsPriority& \%ofNObsPriority($\tau_{var}$,map) \\
                                    &    & map\_priority & \\
                                    & \texttt{GalPlaneFilterMetric} 
        & $M_{f}(f,\alpha,\delta)$ & \% area $\geq$1.0(map)\\
        \texttt{galplaneTimeSamplingMetrics} & \texttt{GalPlaneVisitIntervalsTimescaleMetric} & VIM & \%ofIdealVIM($\tau_{var}$, map)\\
         & \texttt{GalPlaneSeasonGapsTimescaleMetric} & SVGM &\%ofIdealSVGM($\tau_{var}$, map) \\
    \hline
    \end{tabular}
    \caption{Summary of metrics discussed in this work.  The science-based priority maps are inputs to all these metrics, which are calculated using a HEALpix dataSlicer, meaning that the output quantities are calculated per HEALpix. Evaluated metric values are then derived by summing over HEALpix regions for different science cases and timescale categories of variability.}
    \label{tab:metrics}
\end{table}

\subsection{Survey Footprint}
\label{sec:priority_map}
In order to identify the regions of greatest scientific interest to galactic science, we first reviewed the White Papers described in Section~\ref{sec:intro}.  We compiled a list of the survey regions of interest from all papers citing galactic science as a motivation, together with the filter sets and cadences requested.  The sources are summarized in Table~\ref{tab:survey_regions}, including the selection of filters indicated to be most important to each science case.  Many papers placed high value on $g,r,i,z$ observations, particularly those concerned with targets in regions of high extinction or where high cadence in a single filter is preferable, but we note that this should not be interpreted to mean that {\em no} data is required in $u$ or $y$.  Data in these filters are valued, but not always required at the same survey cadence.  This list of regions was extended to include curated catalogs of Open Clusters published in \cite{Kharchenko2013} and Globular Clusters described in \cite{Baumgardt2018}, as well as Star Forming Regions (SFR, \cite{Zucker2020}), which are not adequately represented by a simple function of $N_{\rm{stars}}$/deg$^{2}$.  In addition, we included a survey region optimized for the study of low-mass X-ray binaries (LMXB) from a function which produces probabilistic samples of LMXB positions and distances based on a spatial distribution derived from the galactic mass model \citep{Grimm2002, Dehnen1998}, following the method of \cite{Johnson2019} (Bellm, priv. comm.). 

\begin{center}
\begin{table}[ht]
\caption{Summary of the larger area survey regions included in this study.}
\begin{tabular}{ |c|c|c|c|c| } 
\hline
White Paper & Region & Gal long (l$^{\circ}$) & Gal lat (b$^{\circ}$) & Filters \\
 \hline
 Bono+ (deep survey) & Gal. Plane center & 20 -- +20 & -3 -- +3 & izy \\ 
 Bono+ (shallow survey) & Gal. Plane & 20 -- +20 & -15 -- +10 & ugriyz \\ 
 Gonzalez+ & Gal. Plane center & -15 -- +15 & -10 -- +10 & grizy \\ 
 Street+ & Gal. Plane & -85.0 -- +85.0 & -10.0 -- +10.0 &  griz \\
 Prisinzano+,Bonito+ & Gal. Plane/SFRs & -90.0 -- +90.0 & -5.0 -- +5.0 & gri \\
 Poleski+, Street+ Clementini+ & LMC & 277.8 -- 283.2 & -35.2 -- -30.6 & griz \\
 Street+, Lund+ & Gal. Plane & -85.0 -- +85.0 & -10.0 -- +10.0 &  griz \\ 
 Poleski+, Street+ Clementini+ & SMC & 301.5 -- 304.1 & -45.1 -- -43.6 & griz \\
 Street+(a) & Gal. Bulge & 2.216 & -3.14 & griz\\
 Clementini+ & M54 & 5.60703 & -14.08715 & gri \\
 Clementini+ & Sculptor & 287.5334 & -83.1568 & gri\\
 Clementini+ & Carina & 260.1124 & -22.2235 & gri\\
 Clementini+ & Fornax & 237.1038 & -65.6515 & gri \\
 Clementini+ & Phoenix & 272.1591 & -68.9494 & gri \\
 Clementini+ & Antlia2 & 264.8955 & 11.2479 & gri \\
 Kharchenko+ & Open Clusters & catalog & catalog & \\
 Baumgardt \& Hilker & Globular Clusters & catalog & catalog & \\
 \hline
\end{tabular}
\label{tab:survey_regions}
\end{table}
\end{center}

Table~\ref{tab:survey_regions} shows the extensive overlap between the recommended survey regions, and to first order the superset of these regions could be used as the footprint for a galactic science survey.  

The regions of greatest interest closely follow the sky regions of highest stellar density, owing to the nature of the science concerned.  This varies strongly, particularly across the Plane, and in general the regions of extreme extinction and lower stellar density are of lower (though not zero) scientific priority.  We can therefore refine the survey footprint by prioritizing regions with higher stellar density.  

It is valuable to present the footprint information in the form of a  Hierarchical Equal Area isoLatitude Pixelisation (HEALpix) \citep{gorski2005healpix} map of the sky where each HEALpix is assigned a numerical metric value representing the scientific priority of that HEALpix for galactic science.  Any such metric is an unavoidably crude measure of true value, and does not diminish the importance of downweighted regions.  Instead the purpose of this `priority map' is to indicate the regions where the majority of observations would ideally be directed in order to answer the needs of the science cases considered.  This presents the information in a form which can then be combined with other metrics to evaluate how different LSST OpSims perform for these science cases, and can also be used within the LSST scheduler to determine field priority in observation sequencing.  

We combined the overlapping survey regions in Table~\ref{tab:survey_regions} in the form of a HEALpix map  with  NSIDE=64, resulting in HEALpix sizes of 55.0\,arcmin.  Firstly, we used a map of stellar density in $r$-band derived from the TRILEGAL galactic model \citep{daltio22}, and identified regions of highest interest by selecting sets of HEALpixs with densities greater or equal to 60\%, 70\% and 80\% of the maximum stellar density over the whole sky.  The pixels in these regions were assigned priority weightings of 0.8, 0.9 and 1.0 respectively.  This map highlighted the Galactic Plane, Bulge and Magellanic Clouds. 

We then added to this map the other regions of special interest, such as star forming regions etc, which were initially assigned a priority weight of 1.0. 

We opted to represent the bandpass requirements for different science cases in the following way.  For each science use-case, each of the six filters was assigned a weighting factor to represent the relative frequency with which observations in that filter should be obtained to meet that use-case.  Summed over all passbands, the weights add up to 1.0 for a given use-case.  For example, the filter weighting assigned for fields in the Galactic Plane were {$u$: 0.05, $g$: 0.225, $r$: 0.225, $i$: 0.225, $z$: 0.225, $y$: 0.05}, capturing the lower frequency of observations requested in the extreme-red and -blue bandpasses.  

This enabled us to generate a HEALpix priority map for each filter by iterating over the science cases and assigning a priority for each HEALpix that is effectively the product of the HEALpix priority map and the filter weighting factor, summed over all science cases considered.  In this way, each science case effectively constitutes a `vote' towards the priority of observations of a given HEALpix in a given filter.  

\subsubsection{Galactic Plane pencil-beams}
\label{sec:pencilbeams}
It was apparent from the first iteration of the priority maps that the many science interests in regions around the Galactic center combine to produce a strong weighting towards that part of the sky.  However, surveying {\em only} this region at moderate cadence neglects a key aspect of Galactic Plane science that Rubin Observatory can uniquely excel in, namely its ability to compare populations derived from different regions of the Milky Way.  

A number of science drivers benefit from surveying a wider range of galactic longitudes.
\begin{enumerate}
    \item {\bf Microlensing detection of black holes, stellar and planetary systems: } Surveying a range of galactic longitudes will discover faint or unseen populations  along different lines of sight through the Galaxy.  These lensing events will be caused by different populations in the spiral arms, disk, Bulge or halo depending on the line of sight, and comparing the rates and properties of microlensing events in different fields will produce insights into the population distributions and how they correlate with, e.g. galactic trends in metallicity. 
    \item {\bf Galactic structure: } By surveying across galactic longitudes, LSST will characterize RR~Lyrae and many other indicator variable types, which can be used to more accurately map structures such as the bar as well as resolved stellar populations like Globular Clusters.
    \item {\bf Detection of SFRs, stellar clusters: } LSST co-added images will reach limiting magnitudes of $>$24.9\,mag in all filters, enabling it to detect faint, low-mass populations of galactic clusters and SFRs.  Open Clusters and SFRs are distributed in a wide ($\sim20^{\circ}$) band extending along the Galactic Plane.  
\end{enumerate}

That said, surveying the entire Milky Way Plane within the region $-85^{\circ} \leqslant l \leqslant +85^{\circ}$, $-10^{\circ} \leqslant b \leqslant +10^{\circ}$ at high cadence is likely to require more time than is available.  The region encompasses approximately 354 distinct Rubin pointings, which (assuming typical visit and slew lengths of 42\,s and 30\,s respectively) would take $\sim$7\,hrs to image once per pointing.  To survey this entire region at the nominal cadence of the Wide-Fast-Deep survey would take $\sim$35\% of the available on sky time per year.  This easily exceeds the estimated 10\% of time remaining after Rubin's Science Requirements are met.   

In order to achieve the scientific goals above without compromising the Science Requirements, we modified the priority maps from Section~\ref{sec:priority_map} with a series of narrow `pencil-beam' fields distributed across the Galactic Plane (Table~\ref{tab:pencil-beams}).  Each field would represent a single Rubin pointing, and $\sim$20 such pencil-beams would be clustered around $b\sim0^{\circ}$, but distributed at regular intervals in galactic longitude.  

\begin{figure}[ht]
\centering
\begin{tabular}{cc}
\includegraphics[width=7.5cm]{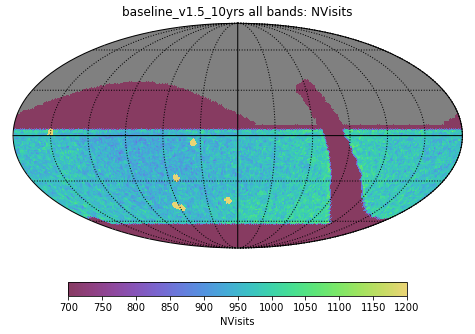} & 
\includegraphics[width=7.5cm]{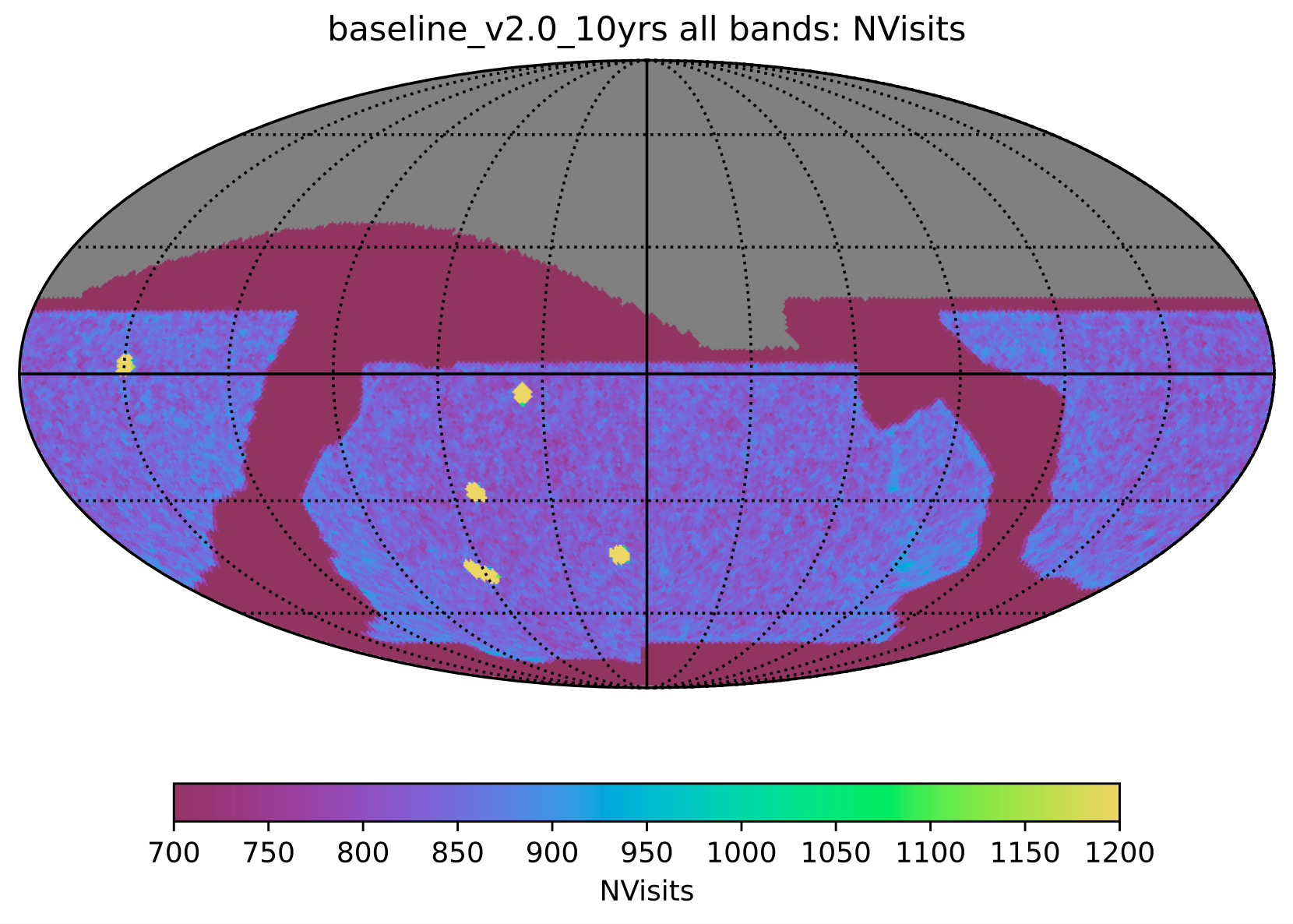} \\
\includegraphics[width=7.5cm]{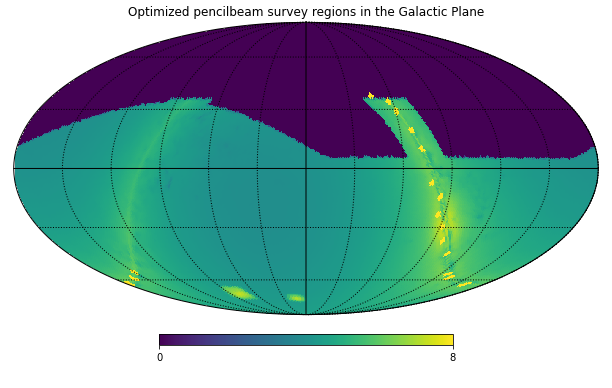} &
\includegraphics[width=7.5cm]{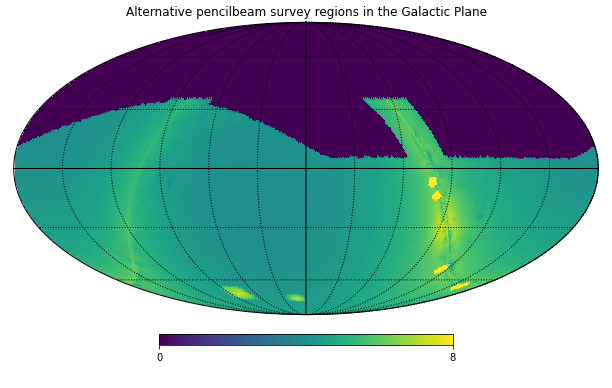} \\
\end{tabular}
\caption{ (Top) Number of visits per sky pointing over all years and filters, derived from the simulated v1.5 (left) and v2.0 (right) survey strategy.  (Bottom row)  Sky maps with a normalized color gradient (values 0--1) indicating the $\log_{10}$(density of stars) based on TRILEGAL models}, overlaid with the two sets of Galactic Plane pencil-beam fields represented in yellow (values $1+$) see Section~\ref{sec:pencilbeams}). Set 1 is plotted on the left, set 2 on the right \label{fig:pencilbeams_map}
\end{figure}

To locate the pencil-beam fields we began with 20 pointings distributed at regular intervals between $-85^{\circ} \leqslant l \leqslant +85^{\circ}$, and $b=0^{\circ}$, and 
then refined them to take into account the uneven extinction across the Galactic Plane.  This was done using maps of the stellar density across the Plane derived from the TRILEGAL galactic model \citep{daltio22}.  Each initial pencil-beam pointing was dithered by $\pm5^{\circ}$ in $l$ and $b$ and optimized by selecting the location which maximized the stellar density for the field.  Figure~\ref{fig:pencilbeams_map} (right panel) shows the optimized locations of the proposed fields. 

The dithering around each survey pointing that is built into the Rubin Observatory scheduler raised a potential issue for the pencil-beam fields.  Each survey pointing is designed to be dithered by up to 0.7$^{\circ}$ \citep{jones2021}.  While this expands the area of a spatially-isolated field and hence the total number of stars included, it also means that not all stars will be observed during every visit.  For spatially contiguous survey regions the dithered pointings will overlap a different area of the survey region, but for a set of spatially isolated fields like the pencil-beams the scientific benefits of including more stars could be overshadowed by the overall drop in cadence.  To explore this potential issue, we also considered an alternative set of pencil-beams fields, which were designed to cover the same total spatial area as the original set, but within only 4 survey regions.  The location of these four regions was based on the EROS2 survey fields towards the Milky Way spiral arms \citep{Moniez2017}.  Figure~\ref{fig:pencilbeams_map} allows the two sets of pencil-beam fields to be compared, and the field locations are given in Table~\ref{tab:pencil-beams}.  It should be noted that since the Galactic Bulge and Magellanic Clouds are considered to be separate survey regions and included independently of the pencil-beams, they are {\em not} part of these field sets.  A comparison of the two sets is discussed further in Section~\ref{sec:refinedDiamond}.  

\begin{center}
\begin{table}[ht]
\caption{Summary of two sets of Galactic Plane pencil-beam fields.  Set 1 fields are single pointings of the Rubin Observatory with a radius of 1.75$^{\circ}$ whereas the set 2 fields have a radius of 3.91$^{\circ}$.}
\begin{tabular}{ |c|c|c|c| } 
\hline
\multicolumn{2}{|c|}{Field set 1} & \multicolumn{2}{|c|}{Field set 2}\\ 
$l[^{\circ}]$ & $b[^{\circ}]$ & $l[^{\circ}]$ & $b[^{\circ}]$ \\
 \hline
280.0 & 0.0 & 306.56 & -1.46\\
287.280702 & 0.0 & 331.09 & -2.42\\
295.394737 & -0.4167 & 18.51 & -2.09\\
306.425439 & -0.4167 & 26.60 & -2.15\\
306.206140 & -0.4167 & & \\
320.153509 & -0.4167 & & \\
324.517544 & -0.4167 & & \\
341.381579 & -0.4167 & & \\
351.578947 & -2.5000 & & \\
0.109649 & -2.0833 & & \\
0.307018 & -2.0833 & & \\
8.421053 & -3.3333 & & \\
17.36842 & -0.4167 & & \\
26.31579 & -2.9167 & & \\
44.01316 & -0.4167 & & \\
44.21053 & -0.4167 & & \\
54.40789 & 0.0 & & \\
66.271930 & -0.4167 & & \\
71.885965 & 0.0 & & \\
80.0 & -5.0 & & \\
 \hline
\end{tabular}
\label{tab:pencil-beams}
\end{table}
\end{center}

\begin{figure*}
\gridline{\fig{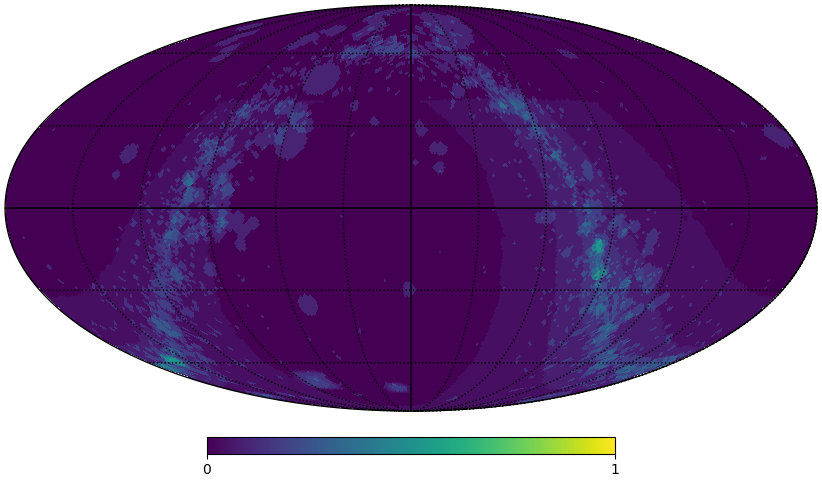}{0.3\textwidth}{$u$-band}
          \fig{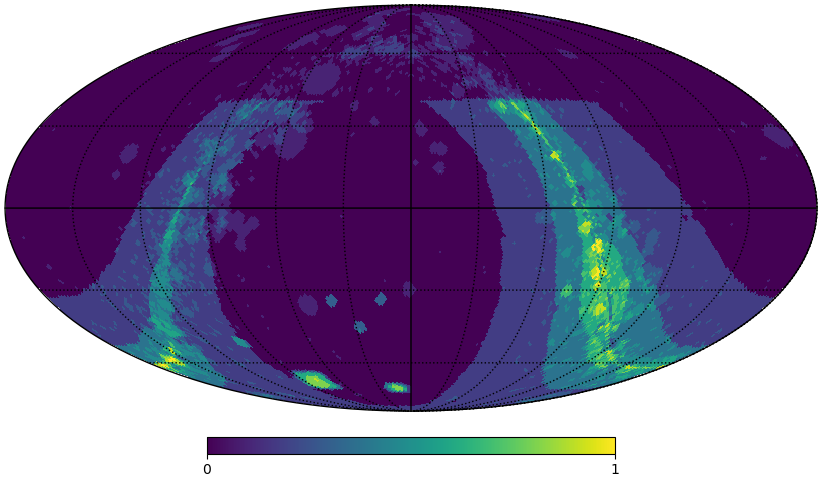}{0.3\textwidth}{$g$-band}
          \fig{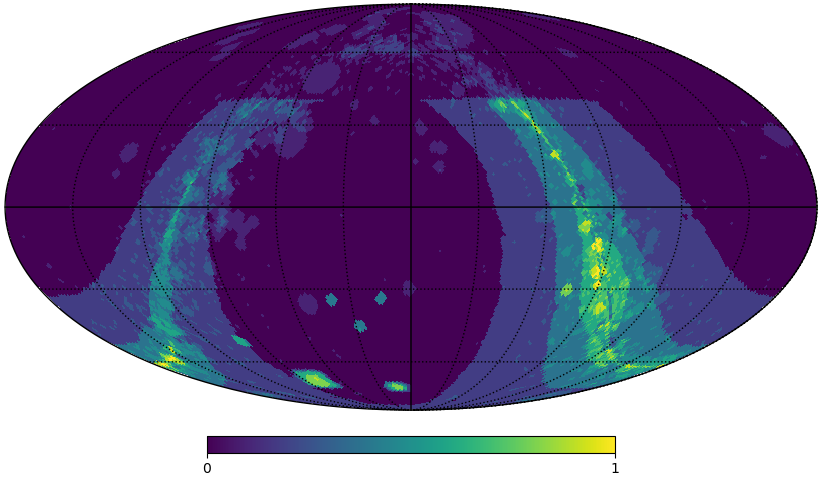}{0.3\textwidth}{$r$-band}
          }
\gridline{\fig{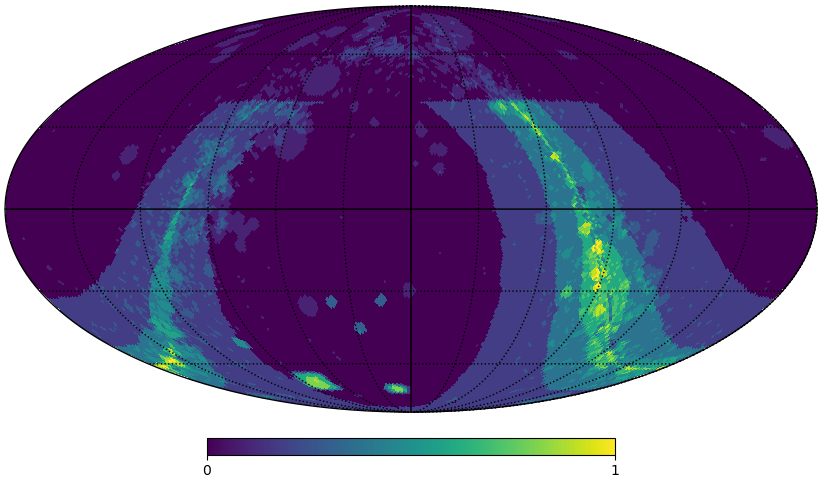}{0.3\textwidth}{$i$-band}
          \fig{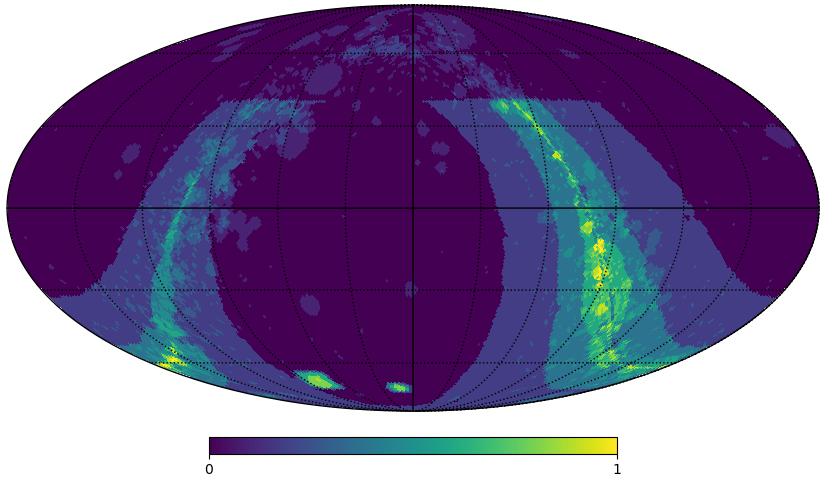}{0.3\textwidth}{$z$-band}
          \fig{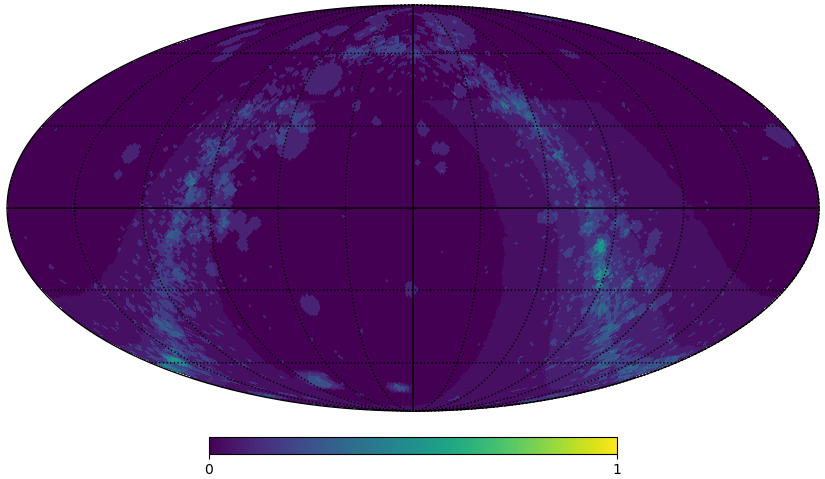}{0.3\textwidth}{$y$-band}
          }
\caption{Sky maps where each HEALpix is weighted according to a priority metric for a broad combination of Galactic Plane science, presented for each filter, and including field set 1 pencil-beams.  \label{fig:priority_maps}}
\end{figure*}

The pencil-beam fields were then included in the combined priority maps described above with filter weightings of {$u$: 0.1, $g$: 0.2, $r$: 0.2, $i$: 0.2, $z$: 0.2, $y$: 0.1}, to ensure that at least a minimum range of fields outside the Galactic Bulge region receive a comparable priority to the central Bulge region.  All pencil-beam fields receive the same priority weighting per filter.  The filter balance chosen reflects the preference for $griz$ for enhanced time-cadence in these fields, but allows for slightly more weight in $u,y$ than the central Bulge, as many of the pencil-beams have less well characterized stellar populations and lower extinction. Although LSSTCam's quantum efficiency is lower in $u$ and $y$ bands, data in these filters is still important to various scientific goals.  $u$-band observations are important for measuring $\log(g)$ for White Dwarfs for example, and for distinguishing hydrogen and helium White Dwarfs, as well as for determining stellar metallicities and as a measure of accretion activity in young stars \citep{bonito2018, Bonito2023}.  This is particularly important for Red Clump stars in the Galactic Bulge \citep{Johnson2020}, despite the high extinction in that region resulting in lower signal-to-noise.  $y$-band data are needed for the detection of Brown Dwarfs, and to characterize instances of their variability due to cloud structures in their atmospheres \citep{street2018bulge, Tan2019}.  The final priority maps with set 1 pencil-beams are shown in Figure~\ref{fig:priority_maps}.
%\& \ref{fig:alt_priority_maps}.  

%\begin{figure*}
%\gridline{\fig{priority_GalPlane_footprint_alt_map_plot_u.png}{0.3\textwidth}{$u$-band}
%          \fig{priority_GalPlane_footprint_alt_map_plot_g.png}{0.3\textwidth}{$g$-band}
%          \fig{priority_GalPlane_footprint_alt_map_plot_r.png}{0.3\textwidth}{$r$-band}
%          }
%\gridline{\fig{priority_GalPlane_footprint_alt_map_plot_i.png}{0.3\textwidth}{$i$-band}
%          \fig{priority_GalPlane_footprint_alt_map_plot_z.png}{0.3\textwidth}{$z$-band}
%          \fig{priority_GalPlane_footprint_alt_map_plot_y.png}{0.3\textwidth}{$y$-band}
%          }
%\caption{Sky maps where each HEALpix is weighted according to a priority metric as in Fig.~\ref{fig:priority_maps}, but including field set 2 pencil-beams. \label{fig:alt_priority_maps}}
%\end{figure*}

The code used to generate these priority maps, as well as the maps themselves in FITS and PNG form, can be found in the TVS Github repository\footnote{ \url{https://github.com/LSST-TVSSC/software\_tools}}. 

\subsubsection{Survey Footprint Metric}
The footprint maps described above are designed to be {\em tunable} in the sense that a larger or smaller survey area can be chosen by setting the HEALpix priority threshold  to lower or higher values respectively.  This allows trade-offs between survey area and cadence to be explored in different simulations.  

To evaluate this in a more quantified manner, the \texttt{GalPlaneFootprintMetric.py} calculates two metrics.  It accepts one of the science priority maps, and a HEALpix dataSlice as input parameters, and selects observations in each filter in turn where exposures reach a minimum 5$\sigma$ depth of ${u: 22.7, g: 24.1, r: 23.7, i: 23.1,z: 22.2,y: 21.4}$\,mag.  These limits were derived from the limiting magnitudes where LSST's catalog of detected sources is expected to be complete to $\geq$50\% in crowded fields \citep{suberlak2018}.  It returns both the number of observations per HEALpix matching these criteria, together with the number of observations multiplied by the priority of the HEALpix in question.  It returns a null value for HEALpixs lying outside the desired survey region of the given priority map for each science case, so a summary metric can be calculated simply by summing over all HEALpixs.  

The metric calculates the number and priority of useful observations of a given HEALpix within the scientifically-desired survey region.  In order to meaningfully interpret these values, we now need to consider what cadence of observation is needed to meet the science needs.  

\subsection{Variability Timescales and Survey Cadence}
\label{sec:timescales}
Time-domain astrophysics, including both transient and repeating variable phenomena, is the science driver behind much, though not all, of the galactic science use-cases considered here.  Accurately capturing all nuances of every type of variability in MAF-style metrics is challenging, and it risks the possibility that the nuance being lost or misunderstood in the interpretation of the results. 

We propose that this goal can be approached by considering the following broad categories of variability according to the typical timescale of the phenomena, e.g. $\tau_{var}$: $<$10\,days, 10--100\,days, 100-365\ and $>$365\,days. These categories can include both transient and (quasi-)periodic phenomena if $\tau_{var}$ is ascribed to either the duration of a transient event or the cycle period, respectively.  Irregular phenomena may be considered to be a sub-class of either transients (if they exhibit extended periods of quiescence between excited states), or quasi-periodic (if the variability is near continuous, $\tau_{var}$ can refer to the average time between high and low states).  For current purposes, the critical aspect is how quickly or how often a variable target changes photometric state, since the survey would need to observe at a minimum of this frequency in order to reliably detect the variability.  Examples of these categories include the following science use-cases.

\begin{itemize}
    \item $<$10\,days, includes exoplanet, white dwarf transits and pulsations, stellar flares, self-lensing by compact objects in binaries, short period stellar binaries, pulsating stars (inc. RR Lyrae, Cepheids), some microlensing (e.g., by free-floating planets, bound planet anomalies, brown dwarfs), variability in young stars, short-term accretion variability.
    \item 10--100\,days, includes most microlensing by galactic stellar-mass lenses, intermediate pulsation periods, short term disk and magnetic instability in Cataclysmic Variables, novae, accretion in young stars,
    \item 100--365\,days, includes microlensing by stellar remnants, long-timescale pulsations, medium timescale disk instability and mass transfer in Cataclysmic Variables, 
    \item $>$365\,days, includes long-period variables, e.g. Miras,  microlensing by black holes and long term disk instability and mass transfer in Cataclysmic Variables.  
\end{itemize}

It should be noted that periodic objects can be detected from low cadence data if the  baseline is long enough, especially in the shortest timescale categories (see for example \citealt{Wiktorowicz2021}).  In terms of survey strategy, this means that the variability timescale does not directly set the minimum cadence requirement.  For these science goals, a better metric is to analyse how well LSST will measure different periods, which is explored in parallel papers including \cite{geller2021}, \cite{dicriscienzo2023} and \cite{prsa2023}.

This work therefore focuses primarily on transient, irregular and quasi-periodic variability.  However, there is an argument in favor of using the following metrics for periodic objects, particularly within the first few years of LSST, to ensure that periodic objects are identified as soon as possible.  In addition to promoting the early study of these variables, it will also improve the identification of transients as they can be excluded from classification.  

Although LSST will generate an alert whenever an object varies by more than 5$\sigma$ from the template image flux, for most of galactic science a single detection is insufficient to reliably identify variability.  
Based on previous follow-up experience \citep[e.g.]{tsapras2019} we require a minimum of 5 observations taken within $\tau_{var}$ to consider a variable target ``detectable'' in the sense of being able to accurately extract targets of interest from the LSST datastream.  It should be noted that this neglects requirements for multi-filter observations within a given timescale, as filter selection is considered with an independent metric described below.  

Applying this requirement to transient phenomena, we derive a maximum-useful observation interval, $\tau_{obs} = \tau_{var} / 5$ for the categories above (\{ \} notation indicates where multiple categories are used to cover a span of timescales):

\begin{itemize}
\item $\tau_{var} < $10 days, $\tau_{obs}$ = 2 days
\item 10 $<$ $\tau_{var} < $100 days, $\tau_{obs}$ = 2 - 20 days; represented as $\tau_{obs}$ = \{5, 11\}\, days
\item 100 $<$ $\tau_{var} < $365 days, $\tau_{obs}$ = 20 - 73 days; represented as $\tau_{obs}$ = \{20, 46.5\}\,days
\item $\tau_{var} > $365 days, $\tau_{obs}$ = 73 days
\end{itemize}

Two values of $\tau_{obs}$ are considered for the $\tau_{var}$ categories between 10 and 365 days, as they encompass a large number of different phenomena of interest which span this timescale range.  We therefore consider six representative categories of variability in what follows.  

\begin{figure}[ht]
\centering
\includegraphics[width=9cm]{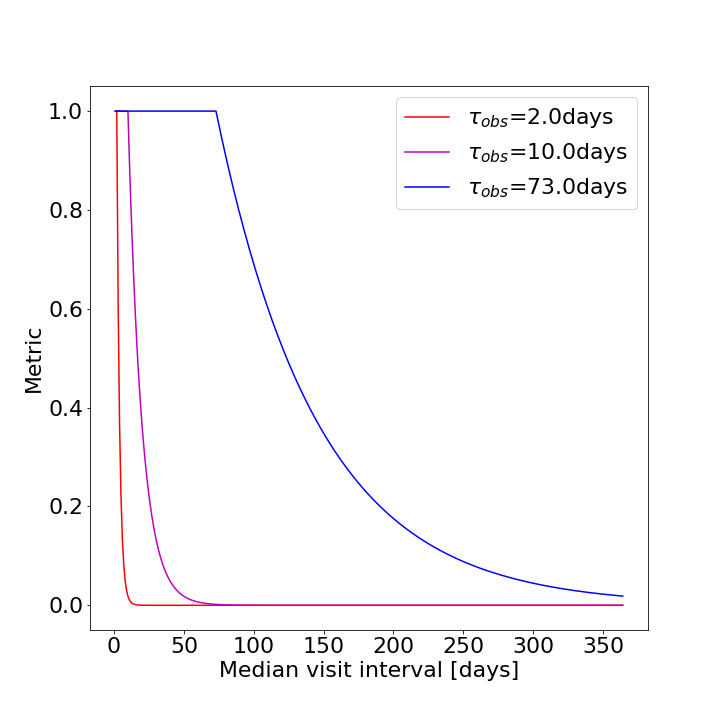} 
\caption{Merit function representing how the scientific value of sequential observations of time-variable objects decreases as the interval between observations exceeds the timescale characteristic of the object's variability.  \label{fig:timescale_function}}
\end{figure}

Typically, observations of time-domain variability, particularly transient phenomena, are most valuable when they occur at the intervals of $\tau_{obs}$.  However, they are usually still valuable even if they occur after this interval, though the value of later observations decreases with increasing interval.  We represent this as a Visit Interval Metric (VIM), a decaying function of the interval between sequential visits (plotted in Figure~\ref{fig:timescale_function}), $\Delta t$ and time $t$, where 

\begin{eqnarray}
  VIM &=& 1.0, t \leq \tau_{obs} \\
  VIM &=& e^{-K(\Delta t - \tau_{obs})}, t > \tau_{obs},  K = 1.0/\tau_{obs}.
\end{eqnarray}

Even if observations are acquired at a regular cadence appropriate to the variability of a given target, there is a second factor impacting whether the target can be characterized from the resulting timeseries.  With the exception of circumpolar objects, a given object is only visible to Rubin for up to a few months each year.  
For variability categories where $\tau_{obs}$ is shorter than the gap between seasons, $\Delta_{gap}$, the season length is not the dominant factor in evaluating a given strategy, and a Season Visibility Gap Metric, SVGM, should return 1.0. But objects with longer variability timescales can be difficult to characterize if the gaps between seasons are comparable to the $\tau_{var}$, so we apply a similar decay function:

\begin{eqnarray}
  SVGM &=& 1.0, t \leq \tau_{obs} \\
  SVGM &=& e^{-K(\Delta_{gap} - \tau_{obs})}, \Delta_{gap} > \tau_{obs},  K = 1.0/\tau_{obs}.
\end{eqnarray}

Consideration of these two components can be used to evaluate how well the timeseries from a given survey strategy will allow variability in these timescale categories to be detected.  

\subsection{Filter Selection}
\label{sec:filter_selection}

The maps of desired survey regions described in Section~\ref{sec:priority_map} were designed to take into account the relative priority placed on observations in different filters by different science cases, by assigning a weighting between 0--1 to each filter for each science case, and then multiplied by the priority of the pointing for each HEALpix.  Relative weights were then summed over all science cases considered to create a priority map in each filter.  These filter-specific priority maps can be used in a metric to evaluate how well a given OpSim meets the filter requirements of the science in the following way.

For each survey strategy simulation, the summed exposure time spent per pointing per filter, $f$, can be calculated as a function of spatial position for each HEALpix as a function of the total exposure time dedicated to that HEALpix, ($R_{exp}$). 

\begin{equation}
    R_{exp}(f,\alpha,\delta) = \frac{\sum_{0}^{Nexp} t_{exp}(f,\alpha,\delta)}{\sum_{0}^{Nexp} t_{exp}(\alpha,\delta)},
\end{equation}
where $f$ indicates the filter bandpass, $\alpha, \delta$ denote the RA and Dec of the HEALpix, $t_{exp}$ refers to exposure time in seconds and $N_{exp}$ gives the number of exposures in each filter per HEALpix.  In the optimum survey strategy, these proportions per HEALpix should correspond to the relative priorities, $p(f,\alpha, \delta)$, of that HEALpix in the priority maps for each filter. This is estimated as a fraction of the total priority for that pixel summed over all filters, 

\begin{equation}
F_{W}(f,\alpha,\delta) = \frac{p(f,\alpha,\delta)}{\sum_{f} p(f,\alpha,\delta)}.  
\end{equation}

The metric therefore outputs $M_{f}(f,\alpha,\delta) = R_{exp}(f,\alpha,\delta)/F_{W}(f,\alpha,\delta)$, and this quantity is evaluated by calculating the percentage of the desired survey HEALpixs where $M_{f}\geq1.0$.

It is informative to evaluate the above metrics for footprint, survey cadence and filter selection individually for the information they provide regarding the strengths and weaknesses of a given survey strategy for different types of variability.  We explored the option of producing a single Figure of Merit which calculated the product of these factors to see if it provided a useful guide survey strategy choices.  However, in practice we found that this inevitably obscures significant nuance for the many different science cases included, so we proceeded with a careful evaluation of the individual component metrics.   

\section{Evaluating the \btz Strategy}
\label{sec:eval_baseline2}

Following community feedback to \cite{jones2021}, the then-nominal survey footprint for the WFD (\texttt{baseline\_v1.5}) was modified to include a region around the Galactic Bulge (\texttt{baseline\_v2.0}).  It is instructive to compare these survey footprints with the galactic science survey footprint from Section~\ref{sec:priority_map}, to quantify how well they would serve the science drivers considered here.  We note that a \bto simulation was also produced, which included the Virgo cluster in the low-dust WFD region and added a requirement to acquire $r$ and $i$-band images in conditions of good seeing (FWMH$_{effective}<0.8$") each year.  Although all simulations were analysed, in practise we found that these changes have minimal impact on the following discussion.  Since \btz continues to be used widely as the canonical reference at the time of writing, we base our comparisons on it.  It should also be noted that each visit to an on-sky pointing consisted of 1$\times$30\,s exposures in \bof compared with 2$\times$15\,s for $g,r,i,z,y$ filter exposures in the \texttt{\_v2.x} simulations.  This means that there are $\sim$9\% more visits available in \texttt{\_v1.5} compared with later implementations, due to the additional overheads.  

For each LSST survey strategy simulation, we identified the region surveyed as the area including all HEALpixs to receive a minimum number of total visits during the 10yr survey.  Observations in all filters are included, but we required that the limiting magnitude reached equal or exceed 21.5\,mag.  This represents the faintest object expected to be detected at 5$\sigma$ in a single LSST exposure, in any filter, of nominal duration of 30\,s in crowded Galactic Plane fields with moderate extinction \citep{suberlak2018}.   While this is a conservative limit in some bandpasses, it was adopted in all filters to simplify the code, because the effect of this constraint is simply to exclude short exposures in a few OpSims designed to explore this option. 
Thresholds for the minimum number of visits were estimated by considering the maximum useful interval between observations, $\tau_{obs}$, described in Section~\ref{sec:timescales}.   We approximated the average seasonal visibility of a given HEALpix from the Rubin Observatory to be 6.5\,months, taking into account the requirement for two visits per pointing per night.  From this we can estimate the expected number of visits to a given HEALpix over the course of the 10\,yr survey if the survey achieved the interval $\tau_{obs}$, for each of the four variability categories.  

\begin{figure}[ht]
\centering
\includegraphics[width=16cm]{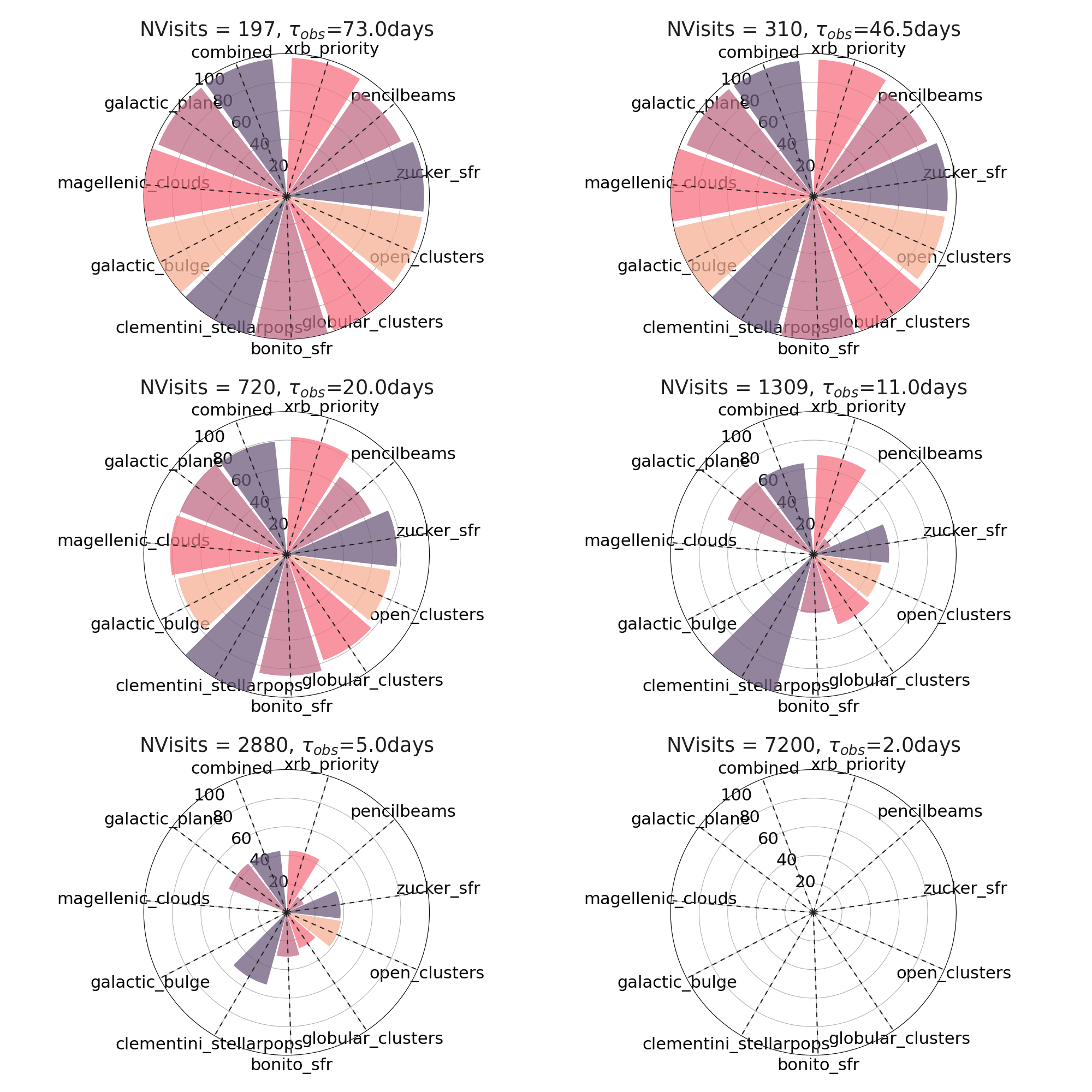} 
\caption{The percentage of the desired survey regions to receive the indicated number of visits over the 10yr survey lifetime in the \bof simulation (\%ofPriority metric), plotted for the four categories of variability timescale. \label{fig:overlap_v1.5}}
\end{figure}

\begin{figure}[ht]
\centering
\includegraphics[width=16cm]{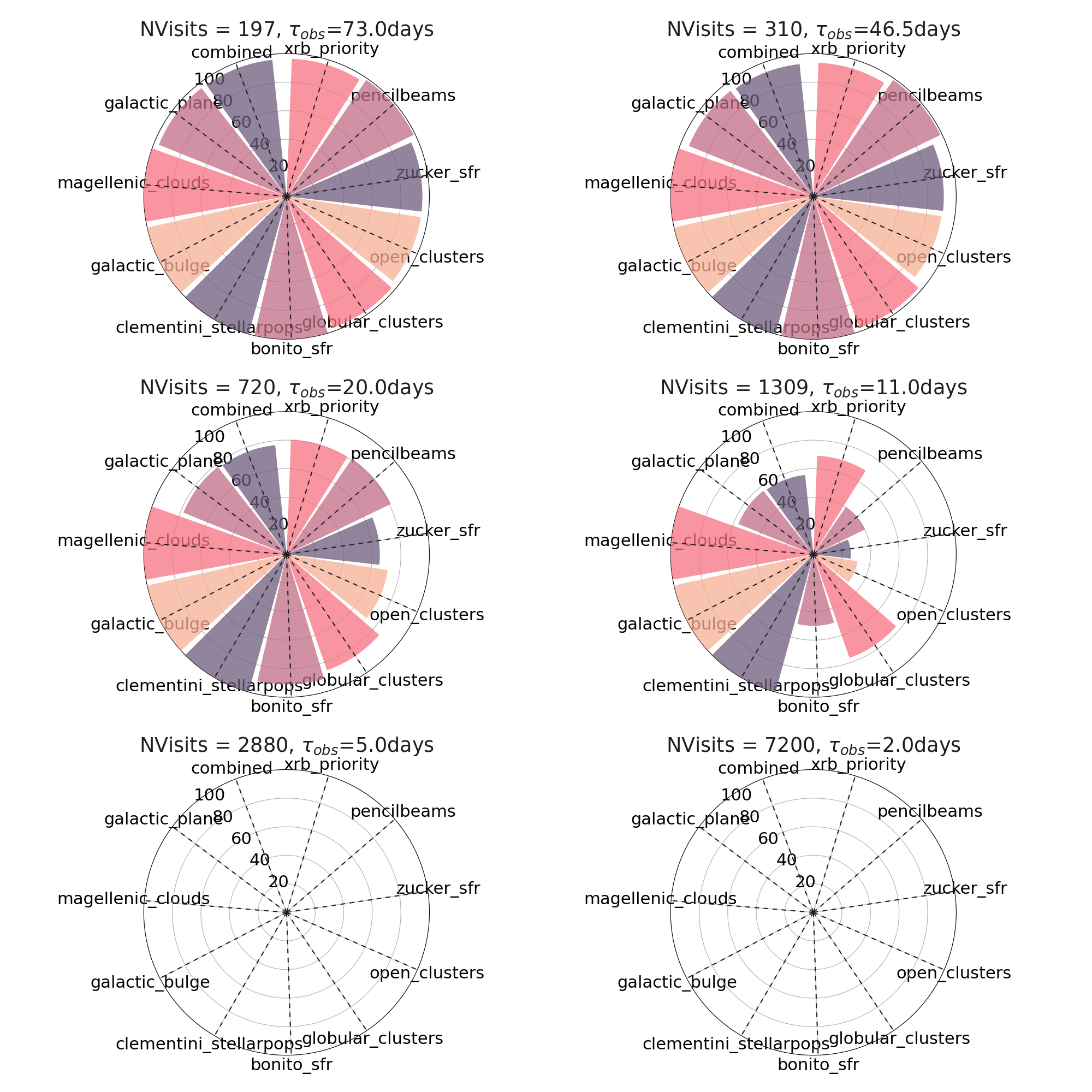} 
\caption{The percentage of the desired survey regions to receive the indicated number of visits over the 10yr survey lifetime in the \btz simulation, plotted for the four categories of variability timescale. \label{fig:overlap_v2.0}}
\end{figure}

The \texttt{GalPlaneFootprintMetric} was used to calculate the percentage of HEALpixs in the desired survey footprint to receive the minimum number of visits in each filter.  We considered both the combined survey region, as well as the survey regions for each science case separately, to gain better insight into the science which most benefits from each survey strategy.  Figures~\ref{fig:overlap_v1.5} \& \ref{fig:overlap_v2.0} summarize the percentage of the desired regions to receive the minimum number of visits needed to characterize the six categories of variability timescale (see Section~\ref{sec:timescales}), for the \bof and \btz simulations respectively.  

Figure~\ref{fig:overlap_v1.5} shows clearly that no region received sufficiently high cadence in \bof to characterize the most rapidly variable targets ($\tau_{obs}$=2\,days / $\tau_{var}<10$\,days) from Rubin data alone.  Notably, the key regions of the Galactic Bulge and Magellanic Clouds did not receive sufficient observations to characterize even intermediate timescale ($\tau_{obs}=5, 11$,d) variables either, although variables in categories with longer timescales are detectable in all regions.  

Strikingly, $\sim$40\% of the combined map of desired regions received sufficient visits to detect variability with $\tau_{obs} > 5$\,d in \bof, while no region did in \btz (Figure~\ref{fig:overlap_v2.0}). This is because \bof dedicated more visits to areas in the Galactic Plane with $5<|b|<15^{\circ}$ whereas \btz concentrated on the central Galactic Plane and Magellanic Clouds.  This means that intermediate timescale variables ($\tau_{obs}\geq11$\,d) are very well recovered from those regions in \btz data.  The results are similar for variables in the intermediate categories for most other regions except Globular clusters, which see an improvement of $\sim$25\%.  {\em The lack of high-cadence data in any Galactic Plane region significantly curtails the detected population of microlensing events, and other transient phenomena, including X-ray Binary/Cataclysmic Variables outbursts. }

\begin{figure}[ht]
\centering
\begin{tabular}{cc}
\includegraphics[width=8cm]{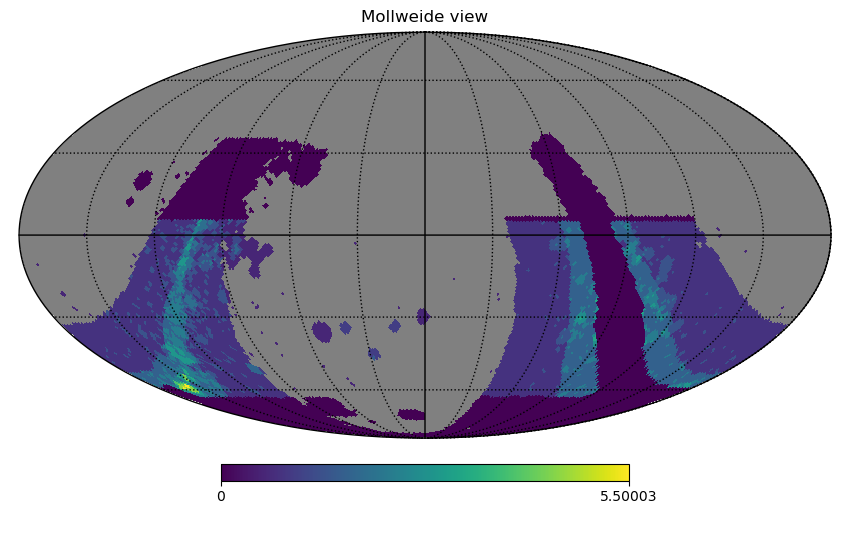} & 
\includegraphics[width=8cm]{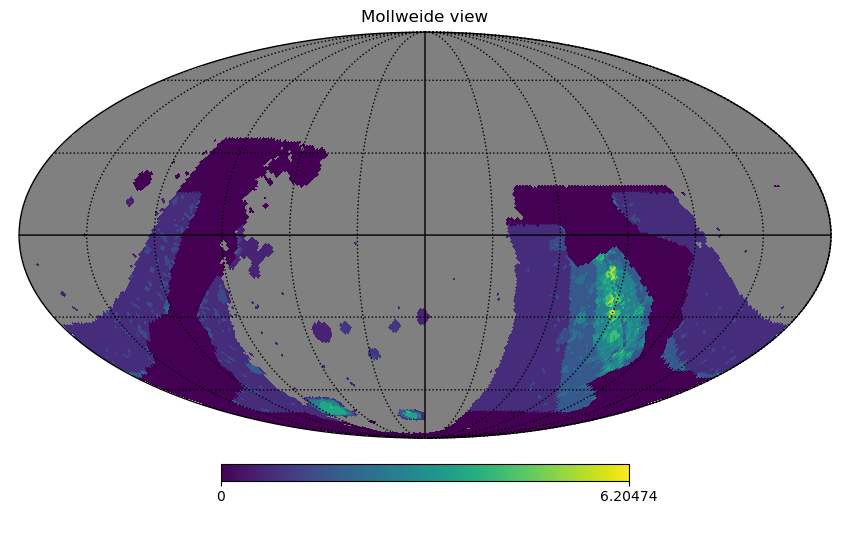} \\
\end{tabular}
\caption{Maps of the combined science priority of each HEALpix regions to receive at least 1309 visits during LSST (average cadence adequate to detect variability on intermediate timescales of $\sim$11\,d).  The dark areas indicate which scientifically desirable regions are insufficiently observed. (Left) \bofns, (right) \btzns.  \label{fig:obs_footprint_priority}}
\end{figure}

\begin{figure}[ht]
\centering
\begin{tabular}{cc}
\includegraphics[width=8.5cm]{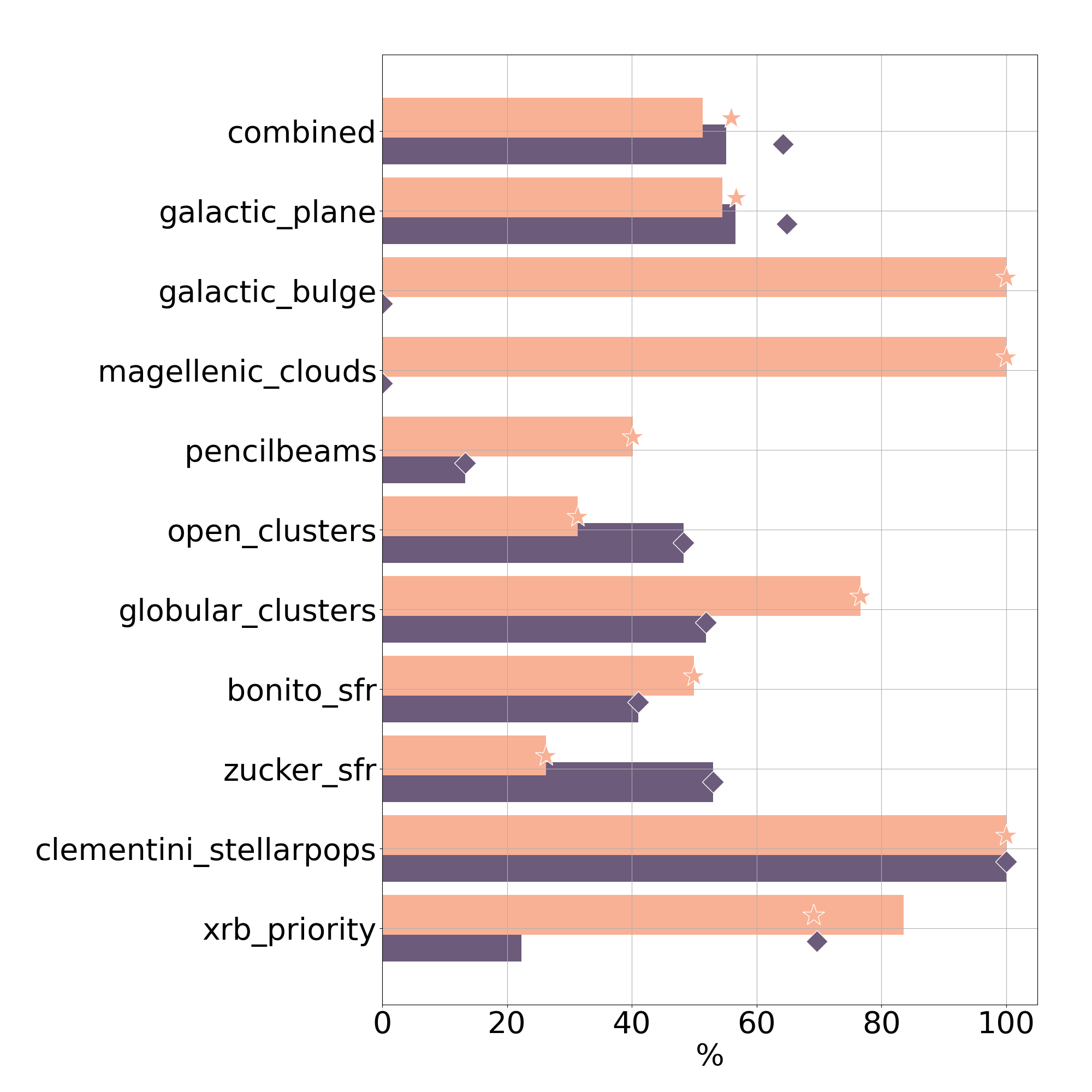} & 
\includegraphics[width=8.5cm]{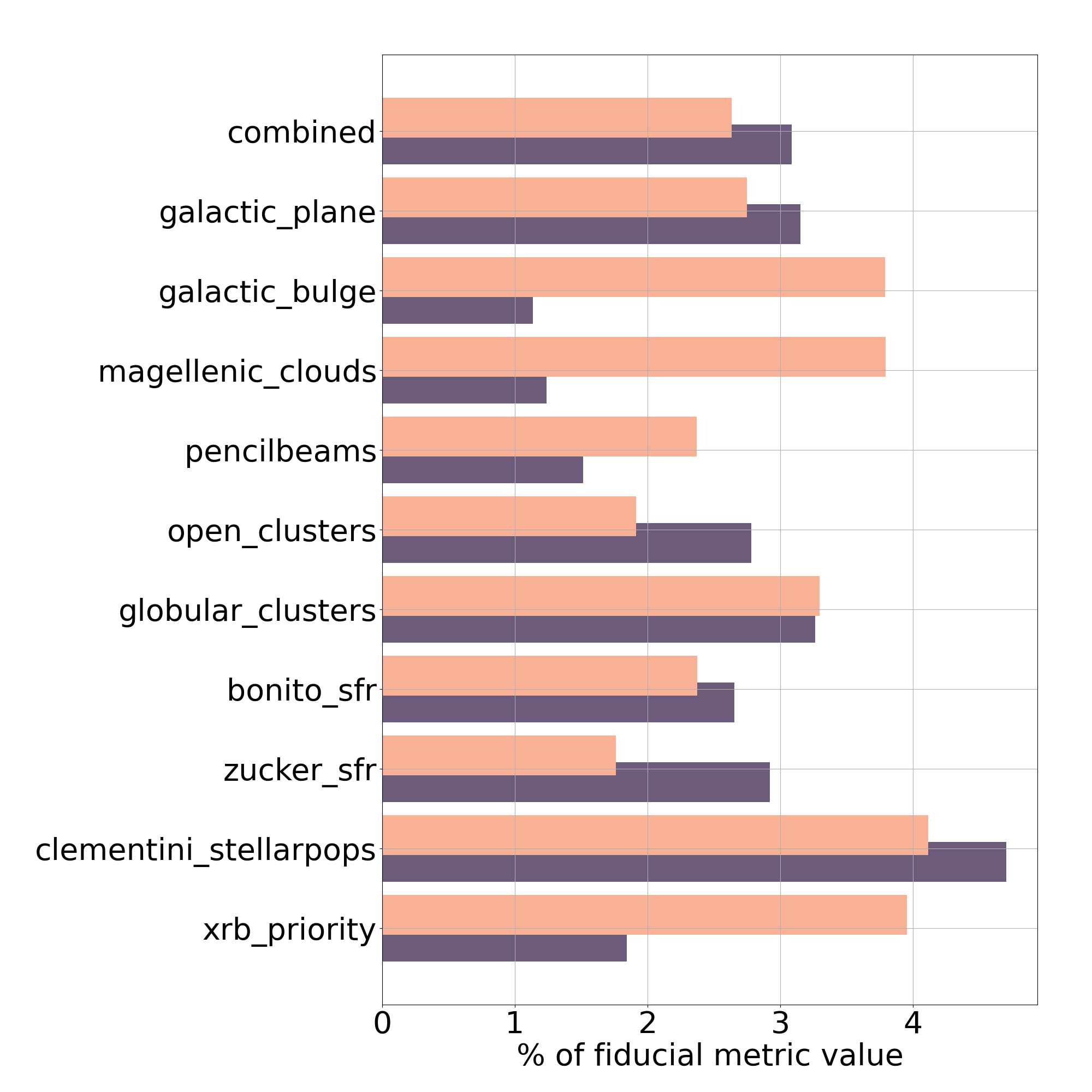} \\
\end{tabular}
\caption{Galactic Plane Figures of Merit used to compare \bof (dark purple, diamonds) and \texttt{\_v2.0} (salmon pink, stars).  (Left) Summed science priority of the HEALpixs within a given science region to receive adequate sampling in the $\tau_{obs}$=11\,d ($\tau_{obs} = 5$\,d) category (bars), compared with the number of HEALpixs sampled (symbols). (Right) The product of the number of visits to a HEALpix and its priority, summed over the survey regions of interest and presented as a percentage of the fiducial value expected from an ideal survey. \label{fig:obs_footprint_FoM}}
\end{figure}

This illustrates one possible trade-off between sky area covered and time cadence achieved, and while the additional visits to the Galactic Bulge and Magellanic Clouds enhance a range of science, we argue that further optimization is possible.  In particular, we note that a great deal of high priority science in the short $\tau_{obs}$ categories is lost when no areas receive visits at high cadence.   This is illustrated in Figure~\ref{fig:obs_footprint_priority}, which presents the regions of the combined HEALpix science priority  map to receive $\geq$1309 visits ($\tau_{obs}>11$\,d) in the \bof and \texttt{\_v2.0} simulations.  While \btz allocates more visits to high priority regions, it still includes more visits in relatively low priority fields at $|b|>10^{\circ}$ instead of the rest of the Galactic Plane at galactic longitudes away from the Bulge. 
The first option for the \texttt{GalacticPlaneMetric}\, ``\%ofPriority'', sums the science priority of all HEALpixs within a given science region to receive sufficient numbers of visits to meet the sampling requirement for each $\tau_{obs}$ category.  This value is presented as a percentage of the summed priority of all HEALpixs within the region in question, giving a measure of whether a given observing strategy focused on the highest priority regions.  We use this to compare \bof and \texttt{\_v2.0} in Figure~\ref{fig:obs_footprint_FoM}, for $\tau_{obs}=11$\,d.  This shows that the revised \btz strategy benefits several science cases with more of their high-priority regions being monitored, notably the Bulge, Magellanic Clouds, Globular Clusters, X-Ray Binaries and the Bonito set of SFRs, and an increase of over 26\% in the monitoring of the pencil-beam fields.  

In contrast, the percentage of HEALpixs from the Galactic Plane region monitored at this cadence actually decreases overall in \btz, by 8.1\%.  But this leads to only a small decrease (2.1\%) in the summed HEALpix priority, reflecting the strategy's focus on the highest priority areas.  

Of greater concern is that the central Bulge Diamond region includes a limited cross-section through the disk and halo populations of the Milky Way, and does not provide good time-cadence of the populations at higher galactic longitudes.  These stellar populations, and their dynamics, can differ significantly, and it is particularly important for microlensing to sample different lines of sight through the galaxy, particularly along the arms of the Milky Way.  
We therefore recommend a less centrally-concentrated Diamond, plus fields in the Magellanic Clouds, with visits distributed across priority regions at a greater range of longitudes. 

We do note significant decreases in the coverage of the Zucker SFR and Open Clusters in \btzns.  This is likely the result of decreased visits to high longitude Galactic Plane fields, relative to \bofns.  

The second GalacticPlaneMetric, ``\%ofNObsPriority'', calculates the product of the number of visits to a given HEALpix and the priority of that pixel, and sums this value over the desired survey regions.  We present this metric (Figure~\ref{fig:obs_footprint_FoM}) as a percentage of the fiducial value that would result from all HEALpixs within a given region receiving sufficient visits to detect variables in the category $\tau_{obs} = 5$\,d.  Though many regions do show an improvement in this metric in \btzns, the Figure of Merit is low for all regions, and decreases overall for the combined science map, despite the increase in visits to the high priority regions.  This reflects the fact that no region is covered at the fiducial cadence in the \btz strategy.  

The goal of these Figures of Merit is to provide a guide to evaluating trade offs between cadence and survey area. If no additional time can be dedicated to surveying the Galactic Plane, we recommend redefining the Diamond region based on the priority maps described above.  We discuss this further in Section~\ref{sec:refinedDiamond}.  

It is important to note that one of the Core Community Surveys that will be undertaken by the Nancy Grace Roman Space Telescope will provide very high cadence (every $\sim$15\,min) NIR imaging of a small region ($\sim$2\,sq.deg, \citealt{Penny2019}) of the central Galactic Bulge.  Owing to Roman's viewing and scheduling constraints, these observations will take place in `seasons' $\sim$62--70\,d long, interspaced with gaps of several months.  Rubin observations are highly complementary to those of Roman in terms of wavelength (optical .vs. NIR 0.48--2.3\,microns), and potentially also in terms of cadence and spatial coverage, if the operational strategies of these contemporaneous surveys are coordinated.  For example, Rubin observations during Roman's inter-season gaps could help to constrain the parameters of microlensing events \citep{street2018bulge}.  An in-depth study of Rubin OpSims specifically focusing on microlensing science is in preparation, and a second paper will explore the benefits of combining Roman and Rubin photometry for these events.  Rubin's survey of a much wider spatial region is particularly valuable as it will encompass a greater range of stellar environments and provide information on the populations of those regions that can then be compared with the results of Roman's in-depth study of Bulge populations.  This is particularly important for inherently rare categories of variables, such as X-ray binaries \citep{Johnson2019}.  We note, however, a few science cases that would not be well served without higher cadence Rubin observations of at least some Galactic Plane regions, such as Cataclysmic Variables (CVs) and Young Stellar Objects.  CVs have blue colors that are not well suited to Roman observations, while Young Stellar Objects exhibit variability on timescales ranging from $<$10\,d -- 100\,d, but are concentrated in SFRs outside the Galactic Bulge. 

\subsection{Survey cadence}

\begin{figure}[ht]
\centering
\begin{tabular}{cc}
\includegraphics[width=7.5cm]{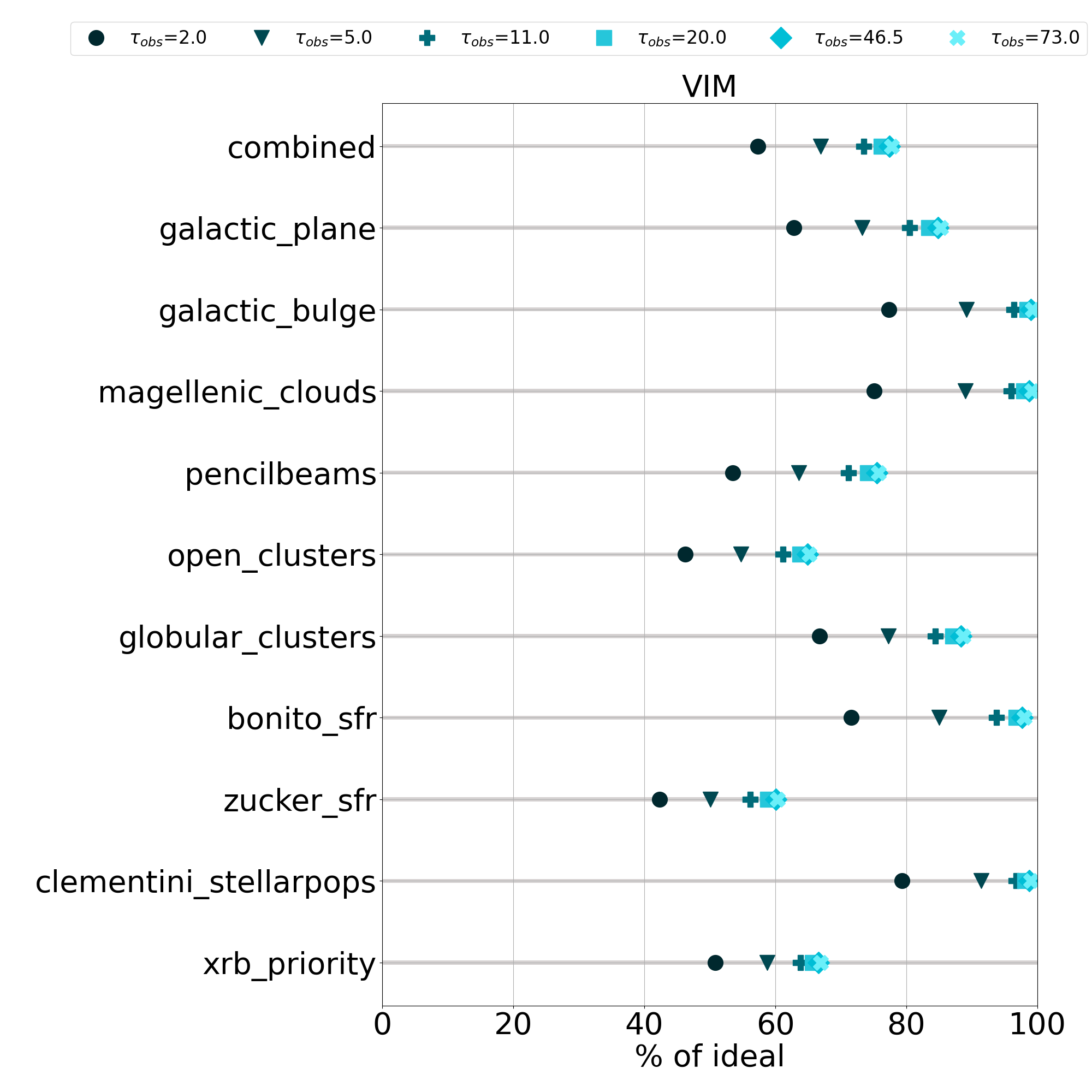} & 
\includegraphics[width=7.5cm]{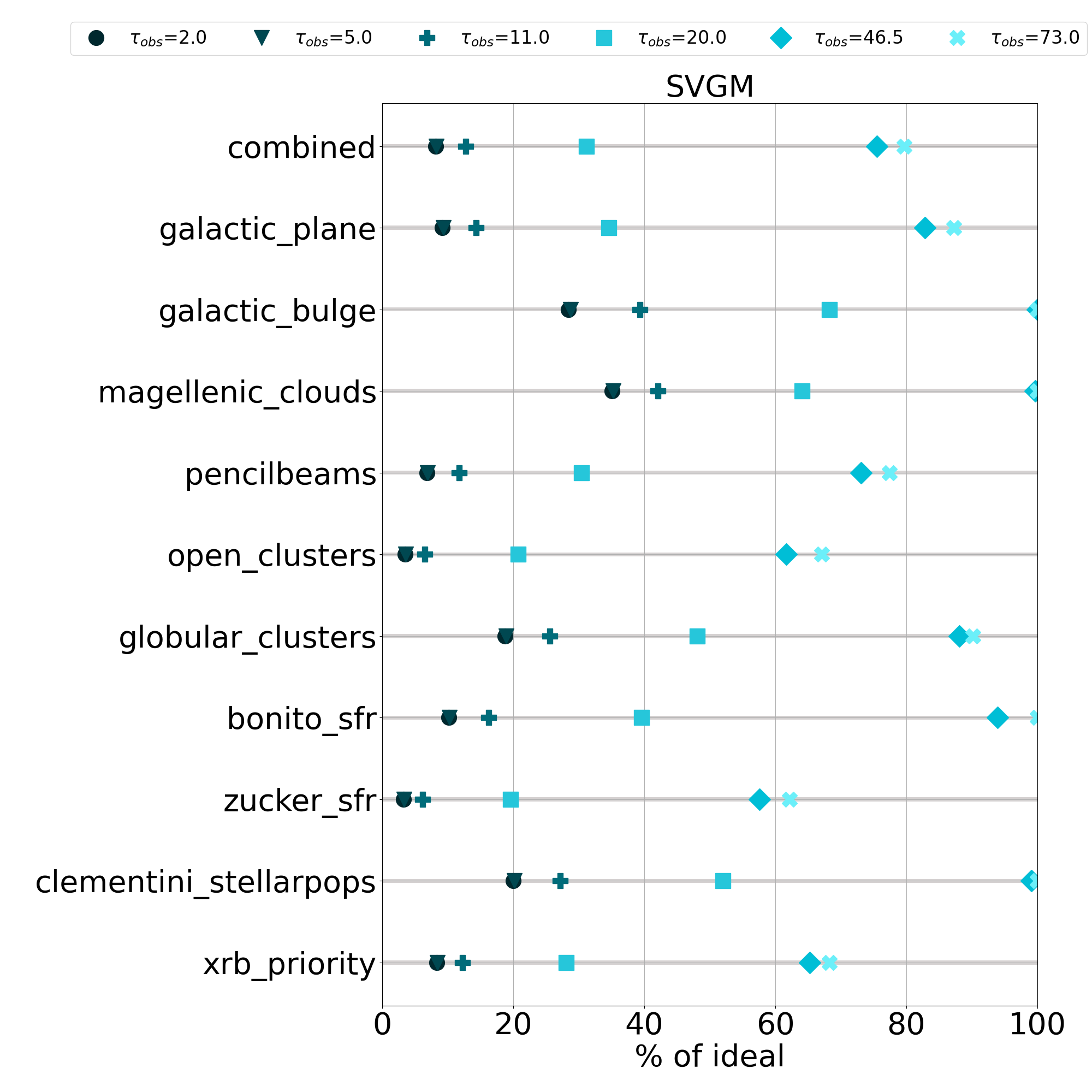} 
\end{tabular}
\caption{Survey cadence metrics evaluated for the \btz strategy. (Left) Survey Visit Interval Metric (VIM) and (right) Season Visit Gap Metric (SVGM). \label{fig:v2.0_cadence_FoM}}
\end{figure}

Figure~\ref{fig:v2.0_cadence_FoM} displays the survey cadence metrics computed for the \btz strategy for all science regions.  In both plots, a value of 100\% indicates ideal cadence over the desired survey region.  It is noteworthy that the VIM metric indicates that the \btz strategy provides good cadence for some of the science regions of interest, particularly the Bulge, Magellanic Clouds and a number of key stellar clusters and SFRs that were not well sampled in earlier strategies.  The lack of monitoring of the regions at higher galactic longitude is reflected in the lower value of the metric for all timescales for the combined, pencil-beams and the X-ray binary science regions. 

All but circumpolar fields will experience some inevitable gaps in survey monitoring due to their annual visibility cycle, and variables (especially transient events) with timescales shorter than this gap will be missed.  The SVGM metric takes the inevitable season gaps into account, and instead evaluates whether the gap is longer than expected (the average expected gap between seasons is 145\,d).  This is primarily of interest for long-timescale phenomena, such as black hole microlensing, for which the lightcurves can be partially sampled if the gap between seasons is elongated.  For this reason, we consider this metric for the longer timescale categories only.  

The \btz strategy maintains relatively uniform sampling across all 10\,yrs of LSST, and thus Figure~\ref{fig:v2.0_cadence_FoM} indicates that long-timescale variables would be characterized in the Plane, Bulge, Magellanic Clouds and in many clusters and SFRs.  The SVGM metric is mainly intended to compare the impacts of rolling cadence implementations, where large latitudinal bands of the sky may be observed intensively in any given year, but then neglected in favor of a different band in the subsequent year(s), potentially leaving long-term variables uncharacterized.  This is discussed further in Section~\ref{sec:rolling_cadences}.

\subsection{Filter Selection}
\begin{figure}[ht]
\centering
\begin{tabular}{cc}
\includegraphics[width=7.5cm]{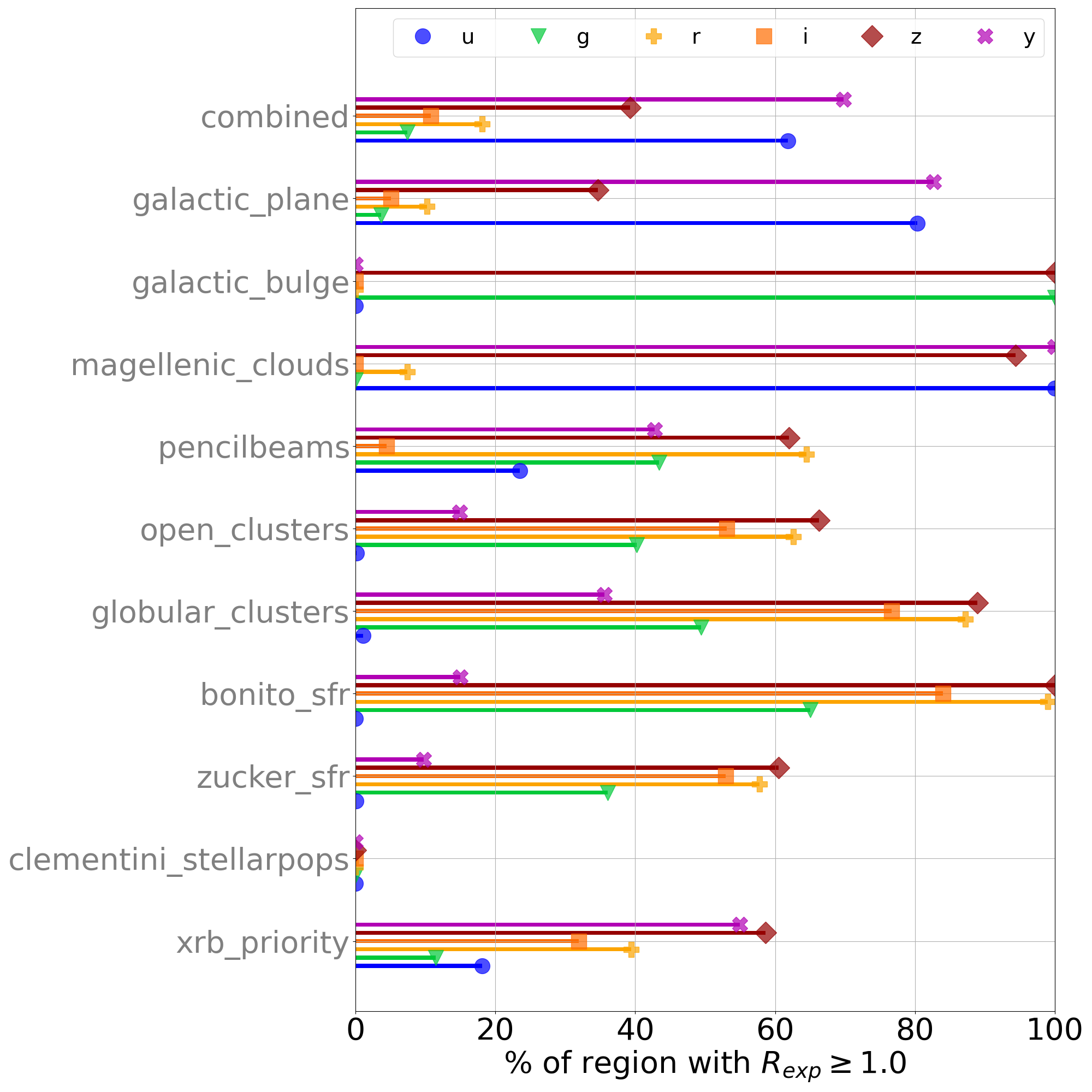} & 
\includegraphics[width=7.5cm]{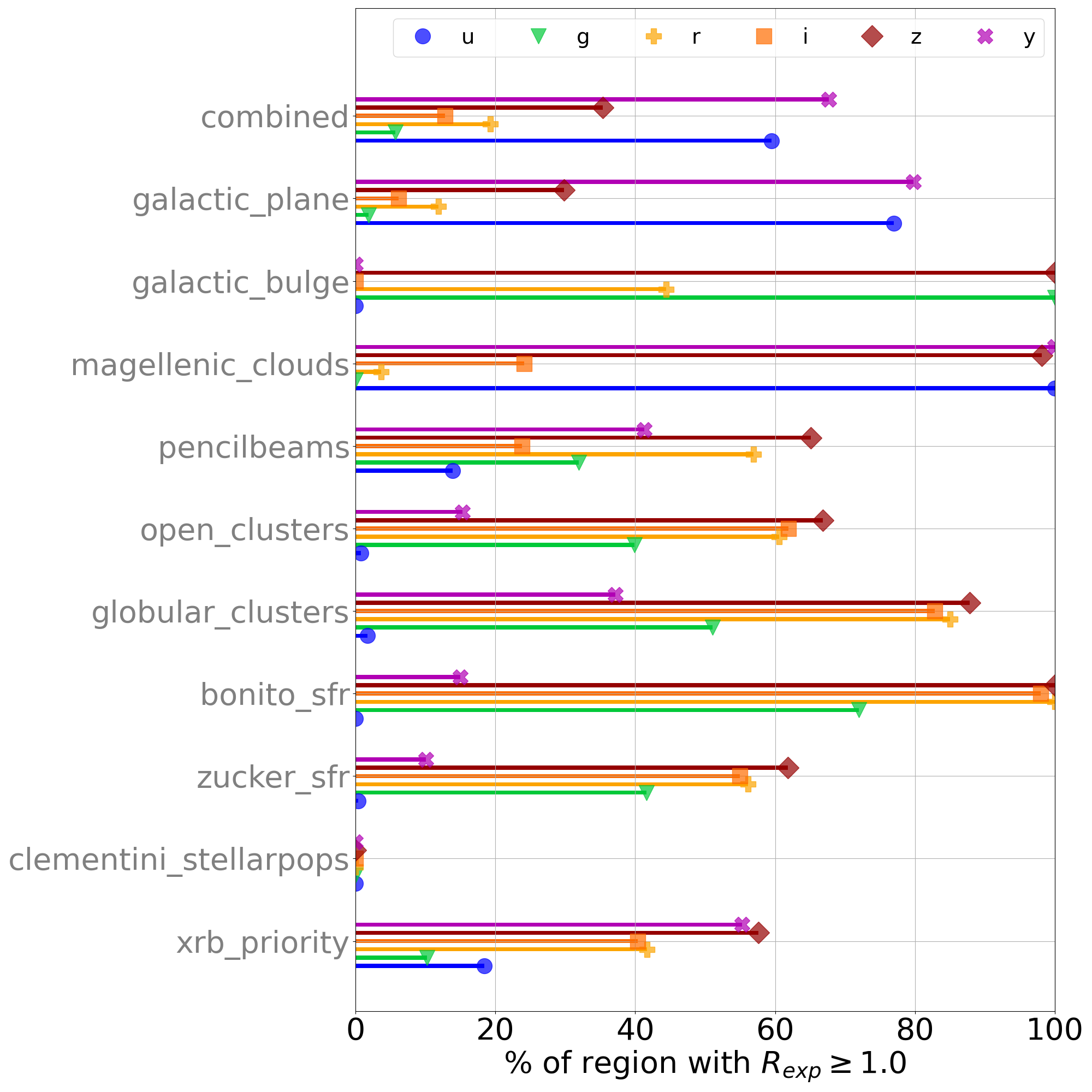}
\end{tabular}
\caption{Percentage of desired survey region to have $R_{exp}\geq1.0$, comparing the time spent dedicated in different passbands for the \btz (left) and \bto (right) strategies. \label{fig:filter_metric_survey_regions}}
\end{figure}

Time-dependent color changes are a key diagnostic feature for many variable types.  To evaluate whether adequate observations were obtained in different passbands, we calculated the percentage of the desired survey region to receive an $R_{exp}$ ratio of $\geq$1.0.  Figure~\ref{fig:filter_metric_survey_regions} presents this metric for the different survey regions and compares the performance of \btz and \bto in different filters.  Since both of these strategies apply a uniform exposure time (30\,s) in all filters, exposure time spent relates directly to the number of exposures acquired in each filter. 

Our analysis indicates that the balance of time spent between different passbands is less than ideal for the combined survey region, for both baseline strategies.  This reflects higher sampling in the $u$ and $y$ filters than is recommended for some regions, particularly those with high-extinction.  Of particular concern is the imbalance between observations in the $g$ band and those in at least one of $r,i$ or $z$, since regular observations in both blue and red bands is required for time-series color observations that will identify and characterize many categories of variable stars. 

Similar concerns are seen for many of the individual science regions, with the Clementini resolved stellar populations being particularly poorly sampled.  

\section{Evaluation of v2.0 and v2.1 Simulations}
\label{sec:eval_v2.0_opsims}
\cite{jones2021} presented a very large set of simulations in `families' designed to explore how varying different parameters and aspects of the survey strategy would impact different science cases.  We evaluated all of the simulations using the metrics described above, and note that this work complements detailed analysis performed for a range of science-specific metrics published in parallel, including the \texttt{MicrolensingMetric} (Abrams et al. in prep) and the \texttt{NYoungStarsMetric} \citep{Prisinzano2022}.  In particular, the recovery of periodic variables is explored in Bonito et al. in prep.  
In this section we summarize our conclusions.  As noted above, the shortest timescales category of variability is not well served in any simulation, so we present results for $\tau_{var}\geq11$\,days.  

\subsection{vary\_gp family of simulations}
\begin{figure}[ht]
\centering
\begin{tabular}{c}
\includegraphics[width=11cm]{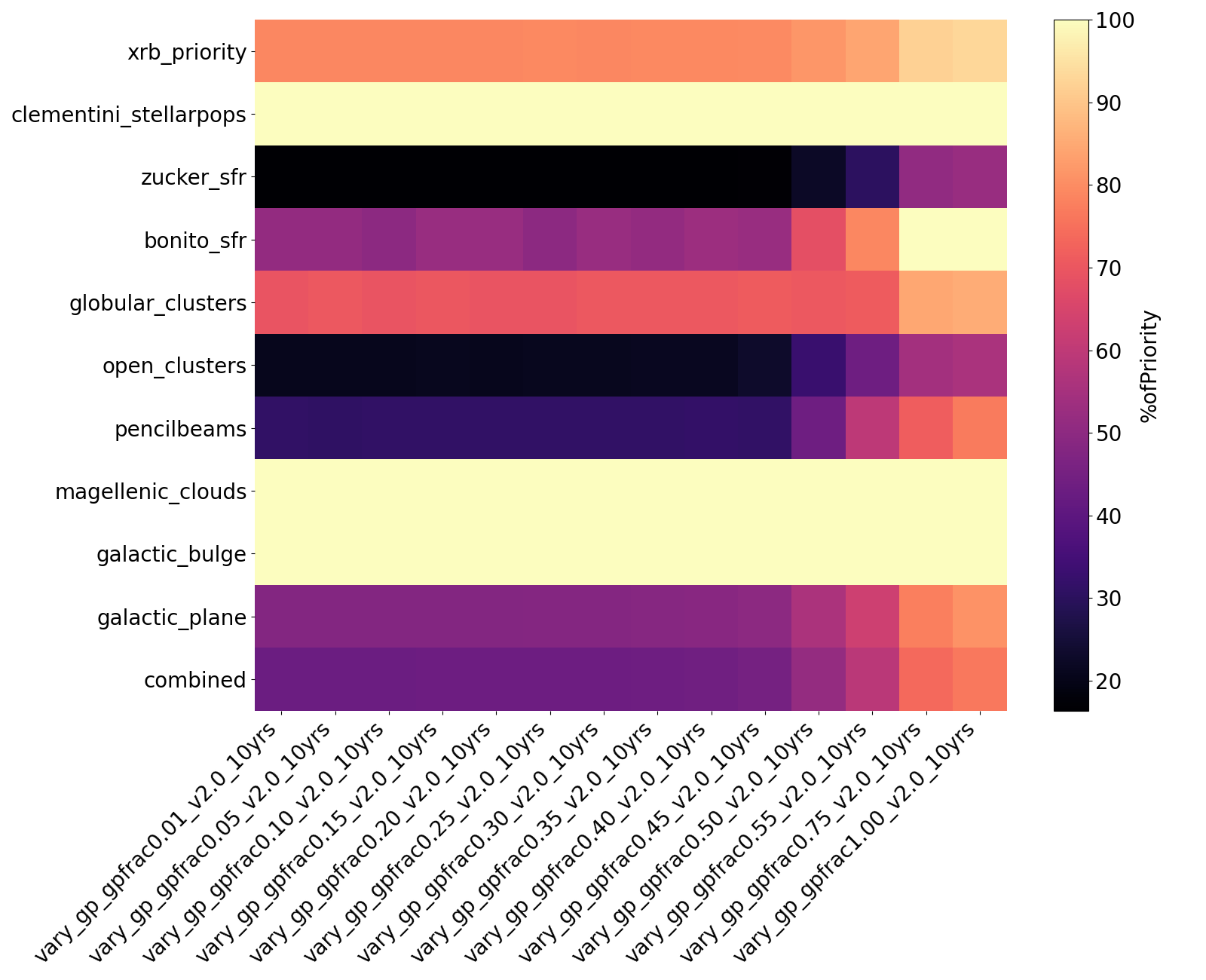} \\
\end{tabular}
\begin{tabular}{cc}
\includegraphics[width=8cm]{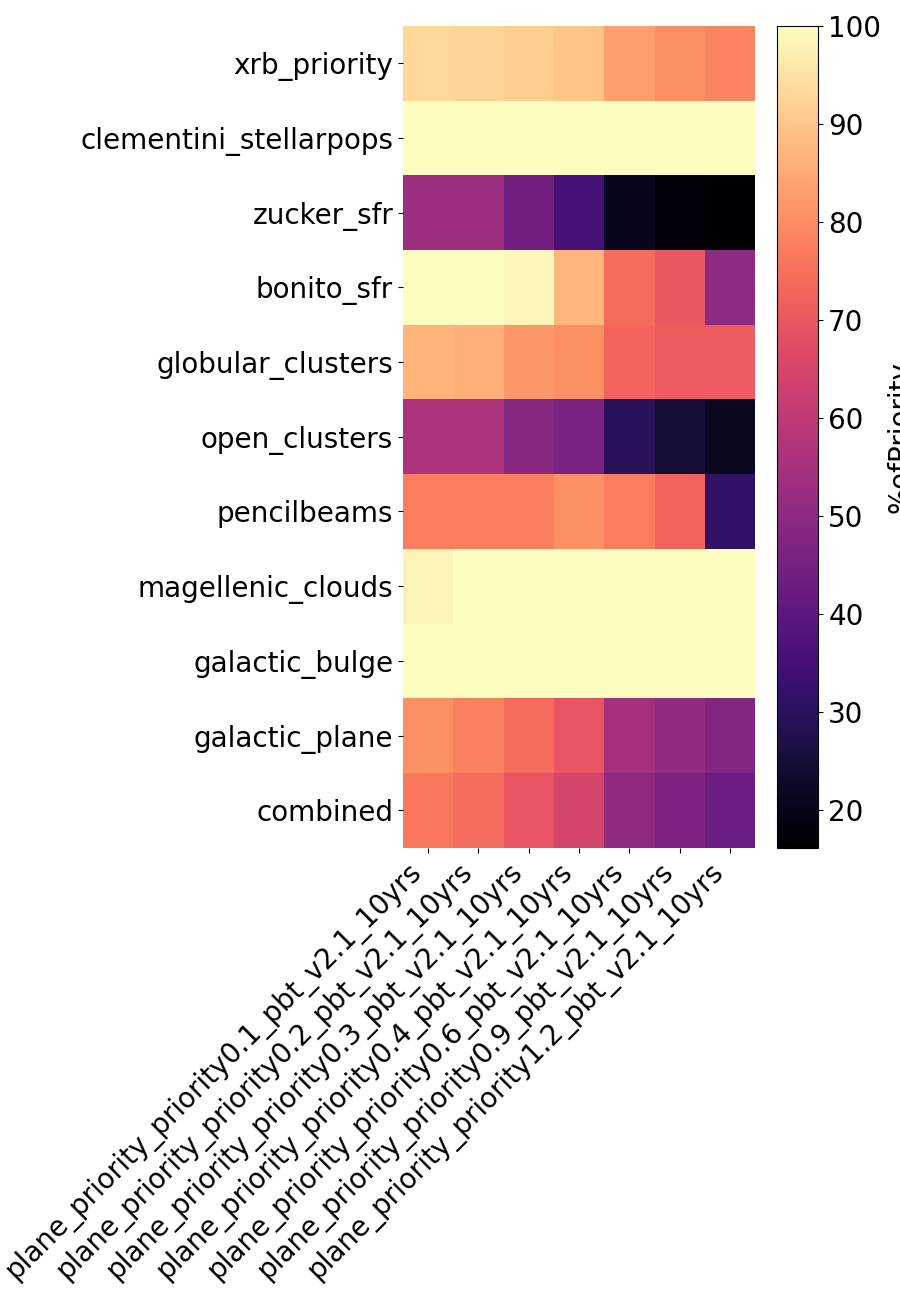} & 
\includegraphics[width=8cm]{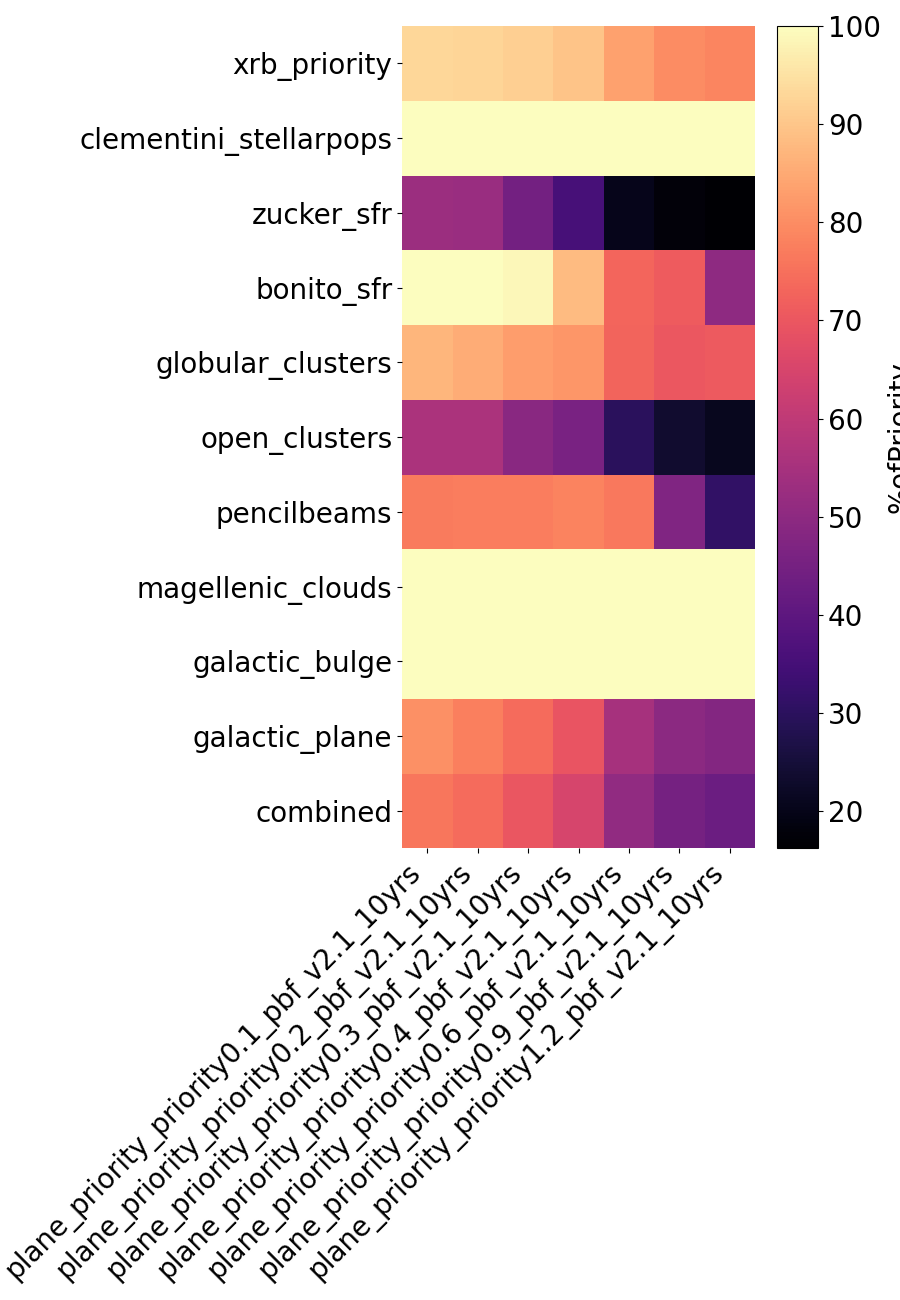} \\
\end{tabular}
\caption{(Top) Heatmap of the \%ofPriority footprint area monitored at sufficient cadence to characterize variability $\tau_{var}=11$\,days, for simulations in the \texttt{vary\_gp} and \texttt{plane\_priority} (bottom) families. In the bottom panels, the left plot summarizes the `pbf' simulations without the pencil-beam fields, while the right plot shows the `pbt' simulations that include them.\label{fig:vary_gp_plane_priority}}
\end{figure}

This family of simulations varied the amount of survey time spent on fields in the region of the Galactic Plane not included in either the Bulge Diamond or the (baseline\_v2.0) WFD region.  Figure~\ref{fig:vary_gp_plane_priority} summarizes the metric data for this family.  

The ``\%ofPriority'' \texttt{GalPlaneFootprintMetric} clearly shows that a greater percentage of the priority area is included when more visits are dedicated to the high galactic longitude regions of the Galactic Plane, particularly for gpfrac weightings $>$0.5, with strong increases in the metric for the combined galactic science region and pencil-beams. Increasing the sampling of regions outside the central Plane increases the total population of variables of many types, including microlensing events, Young Stellar Objects and X-ray Binaries, which boosts the corresponding metric values when summed over the survey region for longer ($\tau_{var}>55$days) timescale categories of variability.  

There is a sweet-spot to be identified in terms of cadence and survey footprint.  Observing a large area with too low a cadence will mean that transient events can’t be properly identified in time for characterization or follow-up.  But reducing the area surveyed in order to boost the cadence will eventually also reduce the total number of transient events discovered.  We recommend evaluating the \texttt{MicrolensingMetric} with two timescales (30d and 200d) in order to identify the optimum balance for transient events (see Abrams et al. in prep).    

\subsection{plane\_priority family of simulations}
This family used the galactic science priority maps described in Section~\ref{sec:priority_map} to select different regions within the Galactic Plane for inclusion in the WFD; effectively exploring the effects of increasing the size of the Diamond region. The `pbf' and `pbt' variants within this family explored the impact of leaving out and including the pencil-beam fields, respectively (see Figure~\ref{fig:vary_gp_plane_priority}, bottom panels).  

Unsurprisingly, the footprint metrics (Fig.~\ref{fig:vary_gp_plane_priority}) show a marked improvement for the combined survey region when the threshold for HEALpix selection is lowered and a larger region is included; this is also particularly important for coverage of SFRs and clusters. 
There is a notable distinction at priority level=0.9 between the maps with and without the pencil-beams included separately.  If a priority selection level of 0.6 or lower is used, then the majority of the pencil-beams regions are sampled at a cadence consistent with the rest of the survey region - regardless of whether they are distinctly included in the map or not.  

We found that adopting a priority threshold of at least 0.4 can be used to select a survey footprint which includes the highest priority regions, while maximizing the cadence achieved. 

In terms of observing known Milky Way star clusters to be used as calibrators for the LSST stellar populations, the changes between the different baselines are minimal: \btz and \bto observe 70 fewer open clusters in the Galactic Plane, and essentially the same numbers of globular clusters. Nevertheless, there are plenty of open and globular clusters left in all baseline plans ($\sim$2000 and 145, respectively), with a good coverage of the age and metallicity space. There are small improvements in the maximum photometric depth reachable in the latest baselines (up to 0.3\,mag in the $r$ band), mostly for clusters in less crowded areas of the Galactic Plane. 

\subsection{Rolling Cadences}
\label{sec:rolling_cadences}
\cite{jones2021} explored a number of different implementations of a rolling cadence strategy, where the sky is divided into separate bands, with some bands being observed more frequently in alternate years, with the `active' bands switching between regions year to year.  It should be noted that the \btz includes a 2-band rolling cadence for the low-dust WFD region.  

Alternatives explored included the use of 2, 3 or 6 bands, each with weightings of 0.5, 0.8 or 0.9, referring to increasing ratio of visits to the active versus inactive bands.  Some realizations also applied the rolling cadence to the central Bulge Diamond, and to the whole sky, while other variations varied in which years the rolling was applied.  

\begin{figure}[ht]
\centering
\begin{tabular}{cc}
\includegraphics[width=7.5cm]{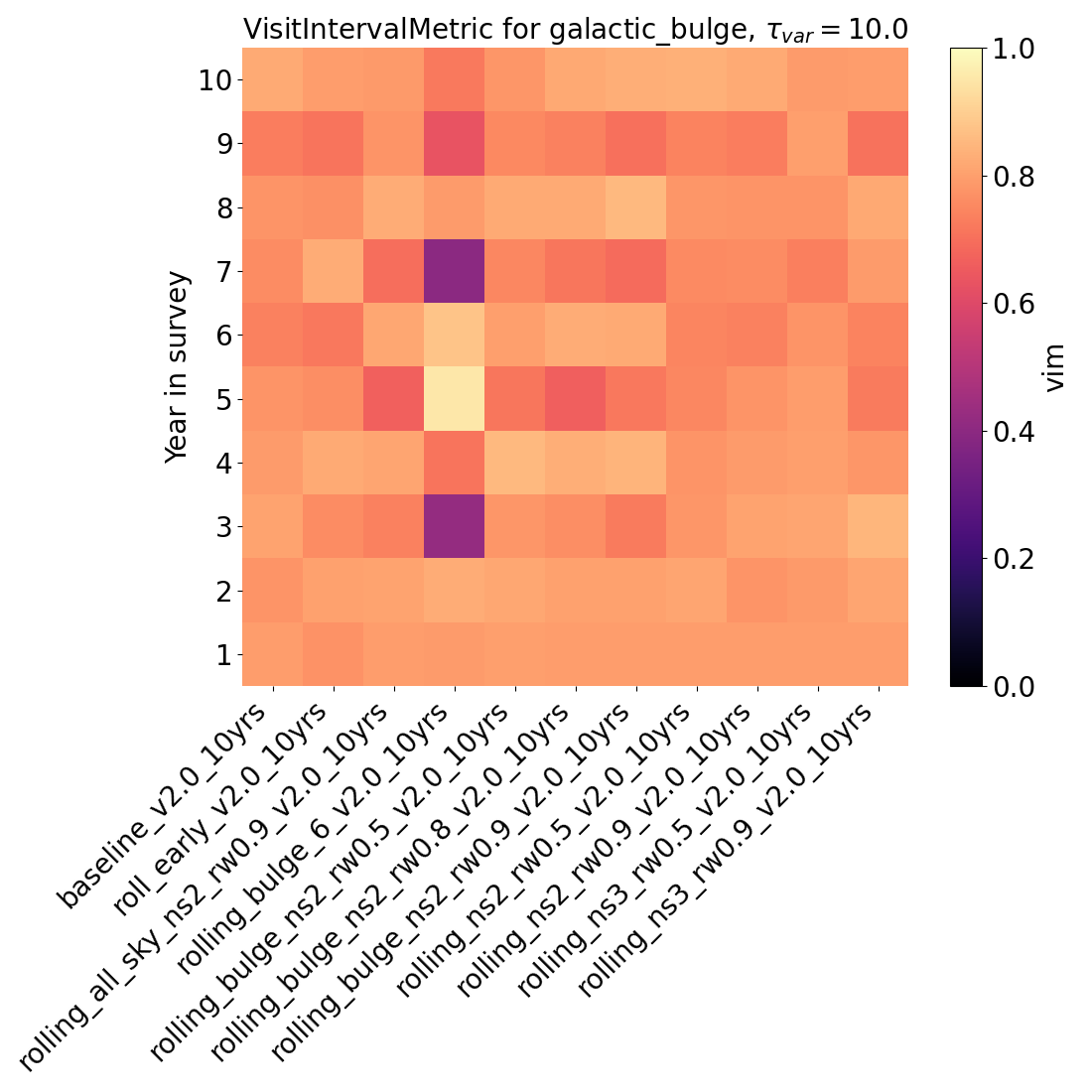} &
\includegraphics[width=7.5cm]{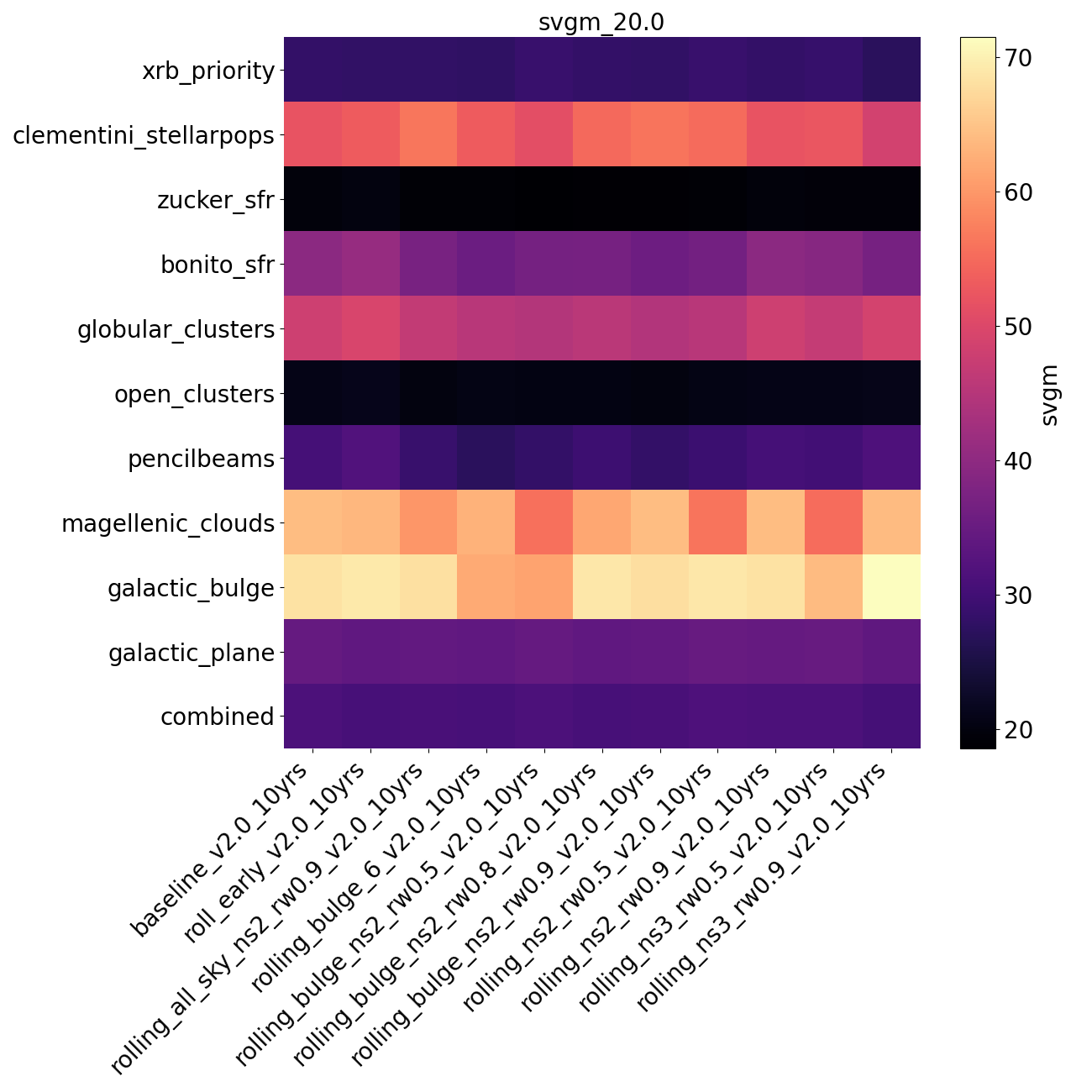}\\
\end{tabular}
\caption{(Left) Heatmap of the \texttt{VisitIntervalMetric} calculated for the Galactic Bulge region for variability $\tau_{var}=10$,days, for a range of rolling cadence strategies. (Right) Heatmap of the Season Visibility Gap Metric calculated for all regions of interest for variability timescales $\tau_{var}=$100\,d. \label{fig:rolling_cadence}}
\end{figure}

Since a similar survey footprint was used in all simulations in this family, we analysed its impact on the cadence achieved for different timescale categories of variability, in the different galactic science regions of interest. In many cases, no impact was seen, because the rolling cadence was not applied to the Galactic Plane.  The exception was the rolling\_bulge family of opsims, and its impact on the detection of short timescale variables ($\tau_{var}=10$\,d) can be seen in Figure~\ref{fig:rolling_cadence}.  

Simulation \texttt{rolling\_bulge\_6} (a 6-band strategy) indicates that this strategy can reach almost the ideal cadence for the shortest timescale (2 day) variability category for 1 or 2\,yrs during the survey, but at the expense of achieving only $\sim$40--60\% of the desired cadence in the remaining years.  

If the years of high cadence in the Galactic Bulge were to coincide with the Roman Mission’s survey of that region, then the complementary datasets would be very valuable for a wide range of science, particularly microlensing planet characterization \citep{street2018bulge}.  This is a high priority science case, but it would come at the expense of lower cadence in the other years, which may be detrimental to Rubin's ability to fill in gaps between the Roman observing seasons.  Regular observations in all years are necessary to ensure that transient events continue to be discovered over a wide area, even if the cadence is so low that characterization depends on additional follow-up. This also ensures that long-timescale phenomena are properly characterized.  {\em We recommend that further exploration of the pros and cons of rolling cadence in the Bulge be undertaken.} This should include comparing simulations of rolling cadence implemented for just the very small Roman Bulge survey region with those of rolling over a larger area, while regular cadence is maintained elsewhere in the Galactic Plane.  At $\sim$2\, sq.deg., this survey region spans just $\sim$20\% of a single Rubin field of view, so a Rubin Deep Drilling Field at this location could not only deliver complementary wavelength and cadence photometry but would also provide data on the wider surrounding region, helping to put the Roman results in context.  

The simulation of rolling cadence across the whole sky is also noteworthy.  In high cadence years, it delivers $\sim$63\% of the desired sampling for the shortest timescale variable categories in the Galactic Plane region.  In ``off'' years, this drops to $\sim$60\%, but the improvement in sampling over the baseline ($\sim$62\%) was small.  
We used the SVGM to explore whether the rolling cadence strategies created any undesirably long gaps between annual seasons of observations of the different regions of interest (Fig.~\ref{fig:rolling_cadence}).  For variability timescales $\tau_{var}\geq$100\,d the metric changed by $<$10\%, suggesting that no gaps in the lightcurves would be detrimental to the characterization of long-timescale variables.  

\subsection{Intra-night Cadences}
A number of LSST White Papers recommended multiple visits per night in various regions, to improve the rejection of spurious transient candidates and the identification of moving objects.  The \texttt{presto} and \texttt{long\_gaps} simulations explored strategies where 2 or 3 visits/night were performed, with different time intervals between them, as well as repeated or different filter selections.  

However, most of these simulations applied these modifications to the WFD survey region.  There was a significant ($\sim$30\%) reduction in the area covered at adequate cadence for $\tau_{var}\geq55$\,days in the Magellanic Clouds if the gap between exposures is short, i.e. $<$3\,hrs.  There was a similar reduction in cadence across Open and Globular Clusters, though the impact was less marked.  Otherwise the results were very similar to the \btz strategy, and shorter timescales of variability are not well sampled. 

Evaluating the balance of time allocated to different filters showed a dramatic variation in the results for different regions of science interest in different filters.  For example, sufficient data were obtained in g-band for the Galactic Bulge in all simulations, but the results in this filter for the Magellanic Clouds, Galactic Plane, X-ray Binaries and resolved stellar populations were very poor.  Conversely, the results in z-band were much better, with most regions receiving at least 60\% of the desired data.

%\begin{figure}[ht]
%\centering
%\begin{tabular}{cc}
%\includegraphics[width=20cm,angle=90.0,origin=c]{figures/percent_100_r_opsim_comparison_heatmap.png} &
%\includegraphics[width=20cm,angle=90.0,origin=c]{figures/percent_100_i_opsim_comparison_heatmap.png} 
%\end{tabular}
%\caption{Heat maps of the percentage of different science regions of interest to receive 100\% of their desired cadence in (left) r-band and (right) i-band in survey simulations exploring intra-night cadence.  \label{fig:intra_night_cadence}}
%\end{figure}

In r-band, the “non-mixed” filter pairings among the presto\_gap strategies generally returned the best combination of filter and cadence, since these focused on the {g,r,i,z} filterset recommended for most galactic science.  The achievable cadence in r for the “mixed” filterset {g,r,i,z,y} was considerably lower in all science regions ($<$20\% of desired).  However, this was reversed in the i-band, where improved cadence was found with the mixed filterset in most presto\_gaps implementations.  

We note that pairs of exposures in different filters are very useful for distinguishing variability classes in real time, and this is particularly important for brokers to be able to identify transients such as microlensing and Cataclysmic Variable outbursts.  The characterization of microlensing will depend on our ability to constrain the source magnification as a function of time in at least two colors during the event.  Regular multi-color observations are also important for the characterization of pulsating stars and other periodic variables.  

{\em We recommend that different filtersets be prioritized with independent cadences separately for the WFD and galactic science regions of interest respectively.  }

\subsection{Other simulations explored}
\cite{jones2021} included a wide range of other simulations designed to explore different scenarios not directly relevant to galactic science.  Nevertheless it was important to assess these simulations as well, to identify any unexpected negative impacts.  Here we briefly summarize our key findings.  

\begin{itemize}
    \item \texttt{Bluer Balance: } Increasing the number of g-band exposures is reflected by an increase in the VIM for Star Forming Regions.  Enhancing both u and g exposures causes a detrimental decrease in cadence for almost all regions.  
    \item \texttt{Long u: } Increasing u-band exposures to 50\,s exposure time enhanced the percentage of some regions of interest that received the desired ratio of visits in u band, provided that the total number of u-band exposures in the survey remained the same.  X-ray binaries, the Galactic Bulge and pencil-beam fields benefited.  However, this came at the expense of the number of observations in the other filters that are important for other science goals.  
    \item \texttt{Vary Exposure Time/Shave: } This strategy slightly reduced the area of the desired footprint covered at a cadence sufficient to detect variability timescales of 55\,d or longer, negatively impacting primarily science from the Galactic Plane, open clusters and star forming regions.  \\
    However, the VIM per year shows significant improvement (1.0 (max) vs. 0.8) over \btz for longer ($\tau_{var}>55$\,d) variability timescales.  This may be due to a higher number of observations reaching the required signal-to-noise thresholds, thereby improving the cadence achieved.  
    \item \texttt{Vary North Ecliptic Spur: } By including more visits to fields at high galactic longitude, more Star Forming Regions are covered at higher cadence.  The simulations with a weighting factor $nesfrac\geq$0.75 show a significant improvement for this science case.  However, it appears to come at the expense of coverage in the Galactic Plane, reducing the cadence in the combined region of interest by $\sim$10\%.  
    \item \texttt{No Repeat: } The imposition of an additional basis function to avoid repeated visits in a single night seems to slightly reduce the number of visits to the Galactic Plane relative to \btons, evidenced by a drop in the area receiving the cadence required in each time category.  This is detrimental to many of our science goals. However, simulations with a higher repeat weight parameter do show a strong improvement in the SeasonGaps metric specifically for the Galactic Bulge region, which is beneficial for long-term variable classes in that region. 
    \item \texttt{Good Seeing: } Requiring that {g,r,i} images be acquired in good seeing conditions for the whole sky results in a small reduction in the areas of interest receiving adequate cadence, relative to \btons.  Given the importance of these images to good image subtraction,  particularly in crowded regions like the Galactic Plane and Magellanic Clouds, we conclude that the benefits outweigh the detriments.   
    \item \texttt{Deep Drilling Fields: } Reducing the number of visits to the DDFs to 3\% causes a small but unsurprising increase in the cadence in galactic science regions of interest, with a corresponding decrease if more time is dedicated to the DDFs.  \\
    Since none of the DDFs lie within the galactic science regions of interest, none of the DDF simulations have any impact on our footprint or cadence metrics.  Some of the DDF ``accordion'' simulations did produce undesirable variations in the frequency of r-band monitoring in the Galactic Bulge specifically, underscoring the importance of implementing different cadences per filter for different survey regions.  
\end{itemize}

\subsection{Microsurveys}
This category of simulations includes a wide range of proposals made in general to support specific science cases, often with specialized constraints on the survey region or cadence.  To some degree, whether adequate on-sky time will be available for these proposals to be included in LSST after the science requirements are met, depends on the hardware overheads relating to slew times, filter changes etc.  Current estimates suggest that some time will be available, but a final decision on the microsurveys will be made after commissioning.  That said, a number of the microsurvey proposals are particularly relevant to galactic science, so we reviewed the simulations provided and summarize our findings below. 

\begin{itemize}
    \item \texttt{Virgo Cluster: } This region lies outside of the desired regions for galactic science, and its inclusion had no impact on the cadence achieved within our desired regions. 
    \item \texttt{SMC movie: } The requested two nights of intensive observations of this relatively small region have no appreciable impact on the metrics for the other regions, while the survey would enable the detection of exoplanets in the SMC, and characterize short-term variables of all kinds.  The alternative strategy proposed (longer-cadence observations over a longer period) would be delivered automatically by including the SMC in the WFD region, as done for \btz and \texttt{\_v2.1}. 
    \item \texttt{Roman: } This strategy proposed a dual cadence: high cadence observations of the Bulge within limited periods where Roman simultaneously observes the same Galactic Bulge Time-Domain Survey field, and lower cadence observations between the Roman survey windows.  Some of the lower-cadence observations would be provided automatically if the Bulge field is included in the WFD region, as in \btzns,\texttt{\_v2.1}.  The VIM shows a small increase for the Bulge region, reflecting the relatively small number of additional observations required for the high-cadence strategy, with no significant impact on other regions.
    \item \texttt{Too\_rate: } Rubin performing 10-50 Target-Of-Opportunity overrides per year appears to have no significant impact on observations in our regions of interest. 
    \item \texttt{North Stripe: } This strategy would increase the number of visits to regions of the Galactic Plane at high longitude, which is beneficial for some galactic science.  
    \item \texttt{Short exposure: } This strategy has no appreciable impact on our metrics, meaning that the additional exposures do not subtract a significant number of visits per region overall.  We note the value of these short exposures for calibrating Rubin's galactic star map at the bright end of its magnitude range, which will allow the stellar populations to be more easily compared between Rubin data and other surveys \citep{Gizis2018}.
    \item \texttt{Multi\_short: } Taking 4 short exposures back-to-back in a single filter per pointing (up to a maximum of 12/year) slightly reduced the cadence achieved in almost all of our desired survey regions, except the Galactic Bulge.
    \item\texttt{Twilight NEO: } These simulations prioritized observations in twilight of areas outside the Galactic Plane where Near Earth Objects are easiest to detect.  This produces a slightly lower cadence (relative to \btzns) in all years for all regions of interest, notably the pencil-beams, Star Forming Regions and the Galactic Plane.  
    \item \texttt{Carina: } These simulations were designed around the \cite{bonito2018} proposal.  The v1.7 simulation represented the best strategy of one visit to the Carina Nebula every 30 minutes, cycling through filters $g, r, i$, and $u$ for 7\,days (with this observation pattern being repeated every 2-3\,yrs).  The goal of the proposal was to execute this pattern for 5 SFRs, including Carina.  Our metrics indicate this strategy significantly enhanced the VIM for the Bonito SFRs as expected (to 95\% for $\tau_{var}$=11\,d compared with 93\% for the \btzns) with no impact to other regions.  A more specialized analysis of the characterization of accreting Young Stellar Objects is presented in \cite{bonito2021}.  
\end{itemize}

\section{Polishing the Diamond: Revised survey region in the Galactic Plane}
\label{sec:refinedDiamond}
%Note that MCs are included in baseline and that's great. Then discuss optimizing the footprint in the GalPlane.
Although the inclusion of the Diamond in the central Galactic Plane and the Magellanic Clouds in the \btz and \texttt{\_v2.1} proved to be a dramatic improvement for many of the science cases considered here, the boundaries of the Diamond were quite approximate in the first round of simulations.  In this section, we describe how the priority maps presented in Section~\ref{sec:priority_map} were used to refine the Diamond region to better reflect areas of high scientific interest.  

Firstly, rather than include all galactic clusters and SFRs in the footprint at the same priority, we implemented a relative priority weighting for the HEALpixs within different regions, based on the recommendations of specialists in those fields of study.  Open Clusters were ranked in priority according to the number of Gaia stars identified within the cluster (Giradi, L., priv. comm.).  SFRs were prioritized according to the order of table~3 in \cite{prisinzano2022b}.  

The maps can be used to directly select larger or smaller regions simply by selecting those HEALpixs with a priority above a lower or higher threshold, respectively.  However, the resulting footprint tends to break into numerous smaller regions that are often separated by gaps larger than the Rubin field of view, unless a very low threshold is used to select a very large area.  This is a natural consequence of the localised nature of some areas of interest (e.g. star clusters) and the non-homogeneous extinction due to patchy dust across the Plane.  

The patchy footprint raised the concern that this would cause the LSST scheduler to spend a higher amount of time slewing between regions rather than exposing, decreasing survey efficiency and the total number of exposures possible over the survey as a whole.  

To explore this possibility, two simulations were compared that implement alternative sets of pencil-beam fields to cover regions of high interest in the Galactic Plane (Table~\ref{tab:pencil-beams}).  The \texttt{pencil\_fs1} simulation added 20 single-pointing pencil-beams, distributed across the Galactic Plane, to the \btz footprint, whereas \texttt{pencil\_fs2} added an alternative set of 4 larger pencil-beams.  Both sets of pencil-beams include the same number of square degrees of sky in total. 

Two metrics were used to evaluate survey efficiency: the number of visits realised to the science regions of interest, and the OpenShutterFractionMetric. We found that larger, consolidated pencil-beams result in a 7\% increase to these regions overall, with little change to the open shutter fraction in pencil-beam fields.  Most of the other regions of interest show small ($\sim$2\%) increases or decreases in their number of visits, depending on whether they overlapped the pencil-beams.  However, coverage of the Bonito Star Forming Regions dropped considerably - receiving 39\% fewer visits - when larger pencil-beams were used.  

In general, the changes in the open shutter fraction were small for all regions, but the standard deviation of the shutter fraction within each science region of interest changed by an order of magnitude for some regions.  Within the pencil-beam regions, the stddev dropped from 0.16 to 0.02 between the smaller and larger sets respectively.  This suggests that the overheads to image different HEALpixs within those regions were more consistent.   

We concluded that survey efficiency can be improved by selecting larger, contiguous regions for galactic science wherever possible.  In order to design this region to maximize the science as fairly as possible across the different science cases, we adopted the following approach.  

The original priority maps (Fig.~\ref{fig:priority_maps}) display numerous spatial regions of HEALpixs (`peaks') at high-priority when plotted in a Mollweide sky projection.  The Mollweide coordinates of the centroids of these regions were identified using the photutils.find\_peaks function.  We consider each peak as a potential pointing for the Rubin Observatory, and identify those pixels which would lie within a single Rubin field of view.  Here a selection threshold was applied, to include only pixels with a priority above a given threshold; this allows us to select larger or smaller survey regions by setting the priority threshold.  We refer to the resulting pixel set as a `cluster'.  

The goal was to identify `superclusters', where the pixel clusters are sufficiently close together, relative to the angular scale of the Rubin field of view, that they can be merged to form a region encompassing several high priority areas.  The superclusters including the greatest cumulative priority value (summed over their HEALpix area) can then be added to a more contiguous, revised footprint region.  

Numerous algorithms exist to identify clusters in parameter space, so we first explored a range of algorithms implemented by the \texttt{sklearn.cluster} package, including DBSCAN \citep{DBSCAN}, MiniBatchKMeans \citep{KMeans}, OPTICS \citep{OPTICS}, MeanShift \citep{MeanShift} and AffinityPropagation \citep{AffinityPropagation}.  We experimented with feature sets including a matrix of the separations between the peaks and their priority values, and tuned the configurable parameters of each algorithm to optimize the results.  Of these, we found that AffinityPropagation and MiniBatchKMeans produced the `best' results.  This was judged by comparing the revised survey footprint produced, with the criteria that it should consist of a small number of contiguous regions, which needed to include the highest-priority peaks and clusters visible on the maps, without also including large areas of low-priority pixels.  

However, this approach emphasized spatial proximity features over the HEALpix priority, which led to a number of smaller (but high priority) regions of the sky being neglected, such as the Small Magellanic Cloud.  We therefore adopted an alternative approach.  We identified potential superclusters where multiple clusters lay within a region $4\times$ the angular width of the Rubin field of view.  The boundaries of each candidate supercluster were defined as ellipses centered on the mean (x,y) of the member clusters with semi-major axes set by the minimum and maximum x,y ranges of the included clusters.  For a candidate supercluster to be accepted, the sum of the priorities of all pixels within the ellipse was required to be greater than the sum of the pixels in the member clusters.  This test prevented superclusters expanding continuously into low-priority regions.  

This approach produced a more consolidated survey footprint, while the use of elliptical superclusters minimized the number of low-priority areas included.  It also allowed us to re-weight the relative priority given to each science case to ensure that science that focuses on specific, small regions was not dominated by science requiring larger areas. 

The original Diamond included a region of $\sim$400\,sq.deg. To identify a more optimized survey region we first used the above process to simply re-distribute this same total area over high-priority regions.  We computed the total area selected by the above process for priority thresholds from 1.0 to 4.4, at intervals of 0.1, and found that a threshold of 2.0 selected a region of 424.7\,sq.deg.

\begin{figure}[ht]
\centering
\begin{tabular}{cc}
\includegraphics[width=8cm]{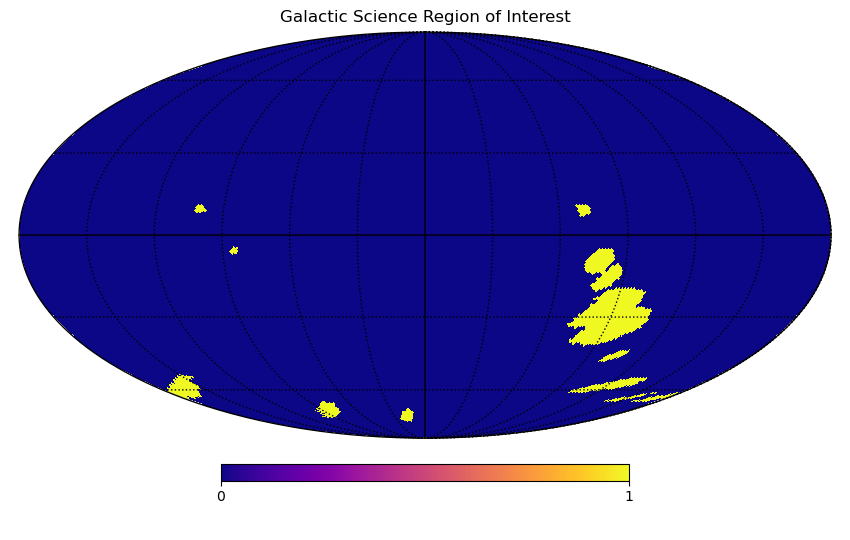} &
\includegraphics[width=8cm]{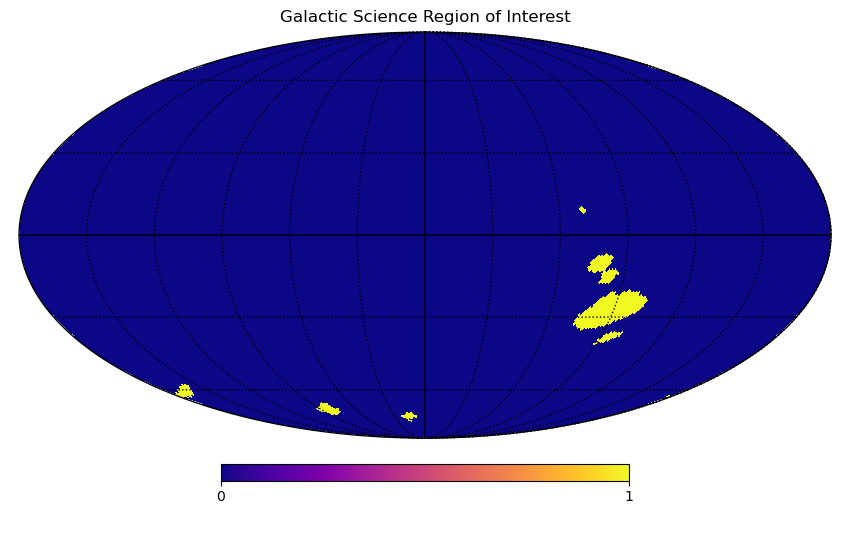} 
\end{tabular}
\caption{Maps of the proposed revised survey region in the Galactic Plane and Magellanic Clouds in Mollweide projection.  (Left) Wider region selected with a projected pixel priority threshold of 1.5, (right) smaller region selected with a threshold of 2.0. \label{fig:aggregated_priority_maps}}
\end{figure}

However, we noted that this region excludes a number of regions of high priority for several science cases, including several pencil-beams and SFRs, so we next identified the smallest survey region that included these areas.  We found that a selection threshold of 1.5 selected an area totaling 937.5\,sq.deg. that included this minimum survey region.  
Both of the resulting survey regions were proposed for the next round of survey strategy simulations, and are shown in Figure~\ref{fig:aggregated_priority_maps}. We stress that these regions represent two possible consolidated survey footprints, but alternative regions can be achieved by adjusting the selection thresholds used.  For example, we recommend exploring the potential benefits of applying rolling cadence to very limited regions, including the Roman Bulge Time-Domain, Magellanic Clouds and selected SFRs, where high-cadence data is most valuable.  By restricting this high cadence region it may be possible to maintain regular cadence over a large area of the rest of the Galactic Plane, thereby balancing the needs of different science cases.  

\section{Conclusions}
\label{sec:conclusions}
We analysed all of the \texttt{\_v2.0} and \texttt{\_v2.1} Rubin survey strategy simulations to evaluate how well they performed for time-domain galactic science, with particular reference to transient phenomena.  While other works have explored the simulations with specific science goals, the objective of this work was to provide high-level metrics to assist with optimizing Rubin's survey strategy for as broad a range of science as possible.  
We approached this by considering the three major observational factors in survey strategy design - sky region, cadence and balance of exposures between filters.  We combined the spatial regions of interest for a range of different galactic science to produce a map indicating the relative priority per HEALpix, and used this to recommend the most scientifically valuable regions of the Galactic Bulge, Plane and Magellanic Clouds to include in LSST.   
A wide range of time-domain galactic science can be represented by categorizing variability according to timescale, but this needs to be considered together with survey footprint, since the spatial distributions of the target objects are often highly localized (e.g. star clusters).  

\subsection{Key Findings and Recommendations}
\btz and \texttt{\_2.1} were shown to deliver significantly improved cadence, relative to \texttt{\_v1.5}, in all of the science cases considered, thanks to the inclusion of a region in the central Galactic Plane and the Magellanic Clouds in the WFD.   However, in our analysis of \texttt{\_v2.0} and \texttt{\_v2.1} simulations, we concluded that very little of the regions of interest to galactic science would receive sufficient visits to reliably detect short ($<10\,d$) timescale variability, unless a rolling cadence strategy is adopted for the Galactic Plane (in whole or a sub-region).  In particular, we note that characterizing short term variability in Young Stellar Objects will require higher cadence in specific regions outside the Galactic Bulge, as discussed in \cite{bonito2021}.  We propose a ``refined Diamond'', consisting of a limited footprint of interest where high cadence observations would be most scientifically valuable, but emphasize that the rest of the Galactic Plane should still receive regular visits throughout the survey.  This ensures that long-term variability can be characterized across the whole Galactic Plane, and that the co-added images will reach sufficiently faint limiting magnitudes to ensure a large volume survey for SFRs and stellar clusters.  We recommend further simulations of a rolling cadence strategy applied to the refined Diamond regions, particularly to coordinate the rolling seasons with the Roman Galactic Bulge Time-Domain Survey, to enhance the characterization of microlensing events from that region.  
Lastly, we note that, for galactic science, there are pros and cons to the choice between 1$\times$30\,s exposures (as adopted in \bof) and 2$\times$15\,s for $g,r,i,z,y$ filters (used in \btz).  Taking fewer, longer exposures results in lower operational overheads, and theoretically allows for $\sim$9\% more visits overall, at least some of which could be used to either enhance the cadence or area covered in the Galactic Plane.  On the other hand, most of the science cases considered here depend on cadence of observation rather than reaching deep limiting magnitudes, and shorter exposures provide more unsaturated photometry of brighter objects.  This allows for the comparison of LSST photometry with that of other surveys, as well as facilitating follow-up observations.

\begin{acknowledgments}
This work was supported by the Preparing for Astrophysics with LSST Program, funded by the Heising Simons Foundation through grant 2021-2975, and administered by Las Cumbres Observatory.  RAS gratefully acknowledges support from the National Science Foundation under grant number 2206828.  This work was authored by employees of Caltech/IPAC under Contract No. 80GSFC21R0032 with the National Aeronautics and Space Administration. YT acknowledges the support of DFG priority program SPP 1992 “Exploring the Diversity of Extrasolar Planets” (TS 356/3-1).
RB acknowledges financial support from the project PRIN-INAF 2019 “Spectroscopically Tracing the Disk Dispersal Evolution.”
RSz acknowedges the support by the Lend\"ulet Program  of the Hungarian Academy of Sciences, project No. LP2018-7/2022.
\end{acknowledgments}


\vspace{5mm}
\facilities{Vera C. Rubin Observatory}


\software{Metrics Analysis Framework, Astropy, pyLIMA}




\bibliography{references}{}
\bibliographystyle{aasjournal}



\end{document}